\documentclass[conference]{IEEEtran}
\PassOptionsToPackage{usenames,dvipsnames,svgnames,table}{xcolor}

\usepackage{nicefrac}
\usepackage{siunitx}
\usepackage{array,framed}
\usepackage{booktabs}
\usepackage{
  color,
  float,
  epsfig,
  wrapfig,
  graphics,
  graphicx,
  subcaption
}
\usepackage{textcomp}
\usepackage{amsmath}
\usepackage{setspace}
\usepackage{latexsym,fancyhdr,url}
\usepackage{enumerate}
\usepackage{enumitem}
\usepackage{algorithm2e}
\usepackage{algpseudocode}
\usepackage{graphics}
\usepackage{xparse} 
\usepackage{xspace}
\usepackage{xcolor}		
\usepackage{multirow}
\usepackage{csvsimple}
\usepackage{balance}
\usepackage{listings}

\colorlet{mgray}{gray!90!black}

\lstdefinestyle{CStyle}{
  language=C,                
  title=\lstname,                 
}

\lstdefinestyle{C++Style}{
  language=C++,                
  title=\lstname,                 
}
\usepackage{lstlangarm}

\lstdefinestyle{arm}{
  language=[ARM]Assembler,        
  title=\lstname,                 
}

\usepackage{
  tikz,
  pgfplots,
  pgfplotstable
}

\usepackage{amssymb}
\usepackage{pifont}
\usepackage{mathtools}

\usepackage[colorlinks]{hyperref}

\newcommand{\binwalk}{Binwalk\xspace}
\newcommand{\vmlinuxtool}{Vmlinux-to-ELF\xspace}

\newcommand{\firmadyne}{FIRMADYNE\xspace}

\newcommand{\vxworks}{VxWorks\xspace}

\newcommand{\hardeningcheck}{Hardening-Check\xspace}
 
\newcommand{\checksec}{Checksec\xspace}

\newcommand{\pwntools}{Pwntools\xspace}

\newcommand{\buildroot}{Buildroot\xspace}

\newcommand{\code}[1]{{\fontfamily{lmtt}\selectfont{#1}}}

\newcommand{\studentfix}[1]{\textcolor{blue}{\textbf{Ruotong: }#1}}
\newcommand{\jun}[1]{\textcolor{red}{\textbf{JX: }#1}}

\newcommand{\para}[1]{\noindent\textit{#1}}

\newcommand{\ie}{i.e.,\xspace}
\newcommand{\eg}{e.g.,\xspace}
\newcommand{\etal}{et al.\xspace}
\newcommand{\etc}{etc.\xspace}
\newcommand{\vs}{v.s.\xspace}

\newcommand{\cmark}{\ding{51}}%

\definecolor{babyblueeyes}{rgb}{0.63, 0.79, 0.95}

\begin{document}
\title{Building Embedded Systems Like It's 1996 \vspace{-1em}}
\vspace{-1.5em}
\author{%
  Ruotong Yu$^{\dagger\gamma}$ \quad Francesca Del Nin$^{\ddagger}$%
  \quad Yuchen Zhang$^\dagger$
  \quad Shan Huang$^\dagger$
  \quad Pallavi Kaliyar$^\S$
  \quad Sarah Zakto$^\P$ \\
  \quad Mauro Conti$^\ddagger*$
  \quad Georgios Portokalidis$^\dagger$%
  \quad Jun Xu$^{\dagger\gamma}$ \\
  $^\dagger$Stevens Institute of Technology
  \quad  $^\ddagger$University of Padua
  \quad $^\S$Norwegian University of Science and Technology\\
  \quad $^\P$Cyber Independent Testing Lab
  \quad $^\gamma$University of Utah
  \quad $^*$Delft University of Technology \\\vspace{-0.5em}
}

\IEEEoverridecommandlockouts
\maketitle


\begin{abstract}
Embedded devices are ubiquitous. However, preliminary evidence 
shows that attack mitigations protecting our desktops/servers/phones 
are missing in embedded devices, posing a significant threat to 
embedded security. To this end, this paper presents an in-depth
study on the adoption of common attack mitigations on embedded 
devices. Precisely, it measures the presence of standard 
mitigations against memory corruptions in over 10k Linux-based 
firmware of deployed embedded devices. 

The study reveals that embedded devices largely omit both 
user-space and kernel-level attack mitigations. The adoption
rates on embedded devices are multiple times lower than their desktop
counterparts. An equally important observation is that the situation
is not improving over time. Without changing the current practices, 
the attack mitigations will remain missing, which may become
a bigger threat in the upcoming IoT era. 

Throughout follow-up analyses, we further inferred a set of
factors possibly contributing to the absence of attack 
mitigations. The exemplary ones include massive reuse of 
non-protected software, lateness in upgrading outdated 
kernels, and restrictions imposed by automated building tools. 
We envision these will turn into insights towards 
improving the adoption of attack mitigations on embedded 
devices in the future.

\end{abstract}

\section{Introduction}
\label{sec:intro}

Embedded devices are running everywhere to connect 
the physical world with the digital world. By estimation, 
there may be up to 35 billion embedded devices installed 
in the wild~\cite{iot-stat-data}. This large-scale deployment 
makes the security of embedded devices critical to our
society. Towards escalating embedded security, it is beneficial 
to gain a systematic understanding of the deficiencies.
Past research has initiated many efforts in this direction~\cite{feng2016scalable,
shirani2018binarm,david2018firmup,costin2014large,chen2021sharing, feng2019understanding,shoshitaishvili2015firmalice,davidson2013fie,cozzi2020tangled,corteggiani2018inception}. 
However, most of them focus on disclosing vulnerabilities in embedded 
devices and understanding the threats imposed
by the vulnerabilities, largely ignoring the other major category of
deficiencies related to the adoption of attack mitigations.
This creates a gap in our understanding.

The gap was gradually realized in recent years, and attempts 
have been made to fill the gap. Earlier research in this line
~\cite{buildsafety} brings preliminary evidence showing a 
lack of adoption of popular mitigations on embedded devices.
More recent studies~\cite{LinuxMIPS,abbasi2019challenges} 
unveil that this lack of mitigations is tied to limited hardware 
or Operating System (OS) support. For instance,
Abbasi~\etal~\cite{abbasi2019challenges} observe that 
deeply embedded devices often lack hardware features
such as Memory Management Unit to enable mainstream 
mitigations against memory corruption exploits. 
These works unquestionably help complete our understanding,
but they (somewhat and unintentionally) leave behind an impression 
that the support-wise barriers are the primary blame for the absence 
of attack mitigations and techniques enabling mitigations 
without those supports (\eg \cite{abbasi2019challenges,
clements2017protecting}) can essentially solve the problem. 
But does this reflect the reality in general?

Aiming to investigate the above doubt, we present 
a large-scale study in this paper. Our angle is to 
look at the adoption of attack mitigations by 
\textbf{embedded devices with all the needed supports},
centering around three dimensions:

\begin{itemize}[leftmargin=0.2in]
    \setlength\itemsep{0.2em}
    \item \textit{With all the needed supports available, do 
    embedded devices adopt the attack mitigations?}
    \item \textit{Is the adoption of the attack mitigations 
    improving over time? Is the upcoming future becoming better?}
    \item \textit{If the attack mitigations are observed absent, 
    what are the possible causes}?
\end{itemize}

\noindent\textbf{Design of Study:} The approach of our study is 
to inspect firmware running on real Linux-based embedded devices, 
seeking to understand their 
adoption of the mitigations listed in Table~\ref{tab:user-space-defense} 
and Table~\ref{tab:kernel-space-defense}. Firmware is targeted to 
match the setup of existing studies of embedded security
~\cite{buildsafety,feng2016scalable,shirani2018binarm,david2018firmup,
costin2014large}. Linux-based devices are considered because
(i) they are typically equipped with high-end hardware, 
which offers modern features needed by the mitigations of interest;
(ii) they represent the dominant type of embedded
devices, according to our data presented in \S\ref{subsec:unpack}. 
The selection of target mitigations is a choice of multiple
factors. First, these mitigations, against the influential 
memory corruption exploits~\cite{realtek,AMNESIA33}, are standard security features 
in common types of computer system (\eg desktops, servers, and 
mobile phones). Second, the mitigations have been integrated
into standard compiling/building toolchains of Linux 
systems, which can be easily deployed. Third, the mitigations 
are released over three years ago. This ensures that the vendors 
have sufficient time to adopt them.

\setlength\tabcolsep{10pt}
\begin{table*}[ht]
  \centering
  \caption{Target attack mitigations in embedded binary programs}
  \vspace{-0.5em}
  \label{tab:user-space-defense}
  \footnotesize
  \begin{tabular}{ c c l c}
    \toprule
    \textbf{Attack Vector} & \textbf{Mitigation} & \textbf{First Release} & \textbf{Default}$^{1}$\\
    \toprule
    Stack Overflow &  Stack Canaries  &    2005 (GCC)  &   \cmark \\\hline
    GOT Hijacking &  Relocation Read-Only  & 2004 (GCC) &   \cmark \\\hline
    Code Injection &  Non-executable Stack  &    2003 (GCC) &   \cmark \\\hline
    Buffer Overflow &  Fortify Source  &    2004 (GLIBC)  &   \cmark \\\hline
    Control-flow Hijacking & Position-Independent (or ASLR-Capable) Code &    2003 (GCC)  & \cmark \\ 
    \bottomrule
    
    \multicolumn{4}{ l }{$^{1}$~Tested on Debian 10
    ``buster,'' released in July 2019, GCC v8.3.0}
  \end{tabular}
\vspace{-1.5em}
\end{table*}
\setlength\tabcolsep{6pt}
\setlength\tabcolsep{2.5pt}
\begin{table*}[ht]
  \centering
  \caption{Target attack mitigations in Linux kernel}
   \vspace{-0.5em}
  \label{tab:kernel-space-defense}
  \scriptsize
  \begin{tabular}{ c c c c c}
    \toprule
     \textbf{Attack Vector} & \textbf{Mitigation} & \textbf{Building Configuration} & \textbf{Release Version} & \textbf{First Release}\\ \toprule
    Stack Overflow &  Stack Protector  &    CONFIG\_HAVE\_CC\_STACKPROTECTOR   &   ARM:v2.6 MIPS:v3.11 PowerPC:4.20 & 2009  \\ \hline
    Privilege Escalation &  PXN$^{2}$   &   --$^{1}$  &   ARM:v3.19 AArch64:v3.7 & 2012 \\\hline  
    Control Flow Hijacking &  KASLR   &  CONFIG\_RANDOMIZE\_BASE   &   ARM:v4.6  MIPS:v4.7 PowerPC:v5.2& 2014 \\ \hline
    Heap Corruption &  Freelist Randomization   &  CONFIG\_SLAB\_FREELIST\_RANDOM   &   v4.7 & 2016  \\  \hline
    Information Leakage &  USERCOPY   &   CONFIG\_HARDENED\_USERCOPY   &   v4.8  & 2016  \\\hline
    Buffer Overflow &  Fortify Source   &   CONFIG\_Fortify\_Source   &   AArch64\&PowerPC:v4.13, ARM-32:v4.17, MIPS:v5.5  & 2017 \\  \hline
    Code Injection &  Non-executable Memory   &  CONFIG\_STRICT\_KERNEL\_RWX   &   ARM:v4.11 PowerPc:v4.13 (MIPS does not support)  & 2017  \\




    \bottomrule
    
    \multicolumn{4}{ l }{$^{1}$~``--'' indicates the mitigation is not affected by the building configuration.}\\
    \multicolumn{4}{ l }{$^{2}$~x86/x64 have similar mitigations called SMEP and SMAP. They are not considered because no x64/x86 kernels are identified in our dataset.}
  \end{tabular}
  \vspace{-1.5em}
\end{table*}
\setlength\tabcolsep{6pt}

Specifically, we collect over 18k firmware images from
38 popular embedded device vendors. Unpacking the firmware
images, we extract nearly 3,000k user-space binaries and 8k 
Linux kernels, as described in Table~\ref{tab:extracted-result}. 
The binaries and Linux kernels are then statically analyzed
to measure the presence of attack mitigations. By breaking 
down the measurement results into different periods, 
we further gain an understanding of the evolution in the adoption
of attack mitigations. Finally, we zoom into the binaries
and kernels to find commonalities that can help
explain the observed absence of attack mitigations.


\noindent\textbf{Results and Findings:} When embedded binaries 
are built, attack mitigations are not frequently adopted.
Considering desktop binaries as the baseline, the overall adoption
rates of embedded binaries are many times lower. For instance,
85.3\% of desktop binaries adopt Stack Canaries, 
but only 29.7\% of embedded binaries do. The lack of mitigations in 
embedded binaries is mainly a ``decision'' of the device vendors, 
except for a few cases where the 
architecture and runtime offer insufficient supports. The analysis 
of kernels presents much worse results. The kernels rarely adopt
attack mitigations. Even the most frequently applied mitigation, 
Stack Protector, only has an adoption rate
of 5.6\%. The absence of kernel-level mitigations is largely 
attributed to one reason. That is, the vendors broadly use older kernels 
where the mitigations are not available, despite newer versions 
supporting the mitigations already exist for a long time.

Further, our evolution analysis identifies no clear 
growth in the adoption of attack mitigations by embedded 
binaries. We hence envision their low rates of adopting 
attack mitigations will
less likely improve in the near future. In contrast, 
we do observe positive changes 
happening to kernels. Older kernels are disappearing
and newer kernels are emerging. As a result, mitigations
such as Stack Protector have been applied more frequently 
in recent years. 

Our last main finding is the following observations to 
help explain why vendors do not apply attack mitiations 
in embedded binaries. 
\begin{enumerate}[leftmargin=0.2in]
    \setlength\itemsep{0.2em}
    \item Vendors of embedded devices often use automated 
    tools to build the systems. The automated tools tend to have 
    a huge delay in importing support of the attack mitigations.
    When an older version of the automated tools is used, 
    which happens in practice, the attack mitigations are often
    not available and thus, cannot be adopted.
    \item A large number of binaries are reused across
    products or even vendors. The lack of mitigations
    in those binaries spreads with their propagation. 
    The vendors cannot change that unless they
    can rebuild the initial binaries.
\end{enumerate}

\noindent\textbf{Contributions:} We make the following contributions.

\begin{itemize}[leftmargin=0.2in]
    \setlength\itemsep{0.2em}
    \item We present an in-depth study to measure the adoption 
    of attack mitigations by embedded devices. The study presents 
    a comprehensive view of the lack of attack mitigations 
    even on platforms that support them. We believe it
    will help raise broader awareness of the threat behind.
    \item We unveil a set of key factors leading to the 
    lack of attack mitigations. These will bring insights
    towards improvement and eventually 
    benefit the security of embedded devices.
    \item We build an update-to-date dataset of Linux-based
    embedded firmware. We create a set of mitigation identification
    tools tailored to embedded binaries and kernels. Both the dataset
    and tools will be made publicly available upon publication
    of the paper. The dataset and intermediate results are 
    released at \url{https://github.com/junxzm1990/iot-security}.
\end{itemize}




\begin{figure*}[ht]
        \centering
        \includegraphics[scale=0.5]{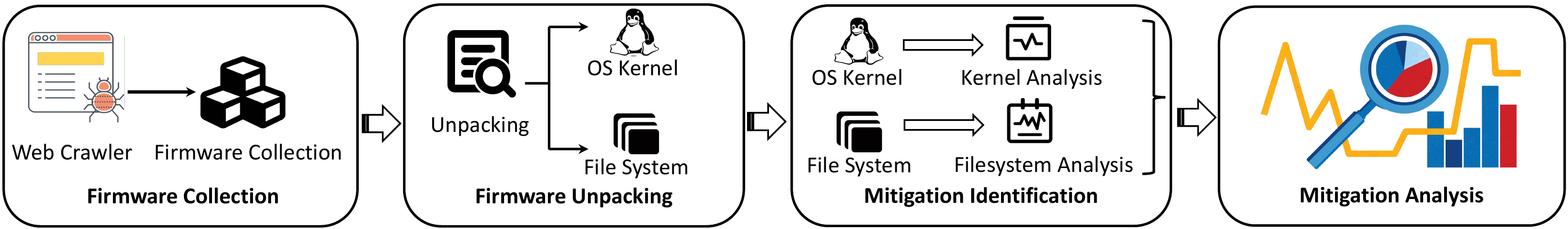}
        \vspace{-0.5em}
        \caption{Workflow of our study}
        \label{fig:workflow}
        \vspace{-1.5em}
\end{figure*}

\section{Challenges}
\label{sec:challenge}

Running our desired study has many challenges. We describe the major
ones in the section.

\para{Building a High-quality Dataset:}
Obtaining a high-quality dataset of firmware images is 
essential but complex. It requires scale, diversity,
and representativeness in the firmware images. Past research 
~\cite{chen2016towards,costin2014large,shirani2018binarm}
has built such datasets. However, we cannot reuse 
them. First, the datasets were collected years ago,
which may not represent what happens at present. Second, 
the public datasets were released in the format
of URL links to the images\footnote{Example: \url{http://firmware.re/usenixsec14/usenixsec14-candidates.yaml.gz}}. 
Most of the links are outdated and invalid today.

\para{Unpacking Firmware Images.} Unpacking the firmware images 
and extracting the required components is also challenging. 
Different vendors organize their firmware images in diverse formats, 
and the vendors typically do not provide information about the composition. 
Past efforts have made plenty of progress in addressing 
the challenge. Tools, such as \binwalk~\cite{binwalk} and 
\firmadyne~\cite{chen2016towards},
can already unpack a broad spectrum of firmware and extract 
individual files like binary programs. However, they are limited 
in identifying and processing kernels. First, kernels in embedded firmware
often have customized signatures, which existing tools cannot capture.
Second, when extracted by existing tools, the kernels are usually 
in the form of raw data, which cannot be further parsed and analyzed. 


\para{Identifying Attack Mitigations:} Binaries in 
embedded firmware are heterogeneous. They run different architectures  
(x86, ARM, MIPS, \etc) and follow various formats (stripped or not, statically 
or dynamically linked, using glibc or uClibc, \etc). The 
heterogeneities affect the identification of mitigations. For instance, 
the identification of Stack Canaries can be done by querying the 
relocation information in dynamically linked binaries, but 
not so for statically linked, stripped binaries. 
Existing tools, including \hardeningcheck~\cite{hardening-check}, 
\checksec~\cite{checksec}, \pwntools~\cite{pwntools}, 
are mainly designed to work in desktop environments. They 
are not aware of these heterogeneities and can present 
reduced utilities when handling embedded binaries. 
Further, most existing tools do not provide presence testing 
of kernel-level mitigations. \checksec~\cite{checksec} offers 
such testing but requires the kernel is booted and running,
which cannot scale to support a large-scale study like ours. 
New tools to statically identify mitigations from kernels
are needed. 

In the following two sections, we detail how we overcome 
the above challenges, 
following the workflow in Fig.~\ref{fig:workflow}.

\section{Data Collection and Processing}
\label{sec:data}




\subsection{Collecting Firmware Images}
\label{subsec:collect}

Our study starts with preparing web crawlers to scan the websites 
of mainstream embedded vendors and collect their firmware images. 
Such crawlers have been developed by past 
efforts~\cite{chen2016towards}. However, due to the dynamic nature of vendors' websites, 
they yield poor results \textit{today}. We update and extend the crawler released 
at~\cite{firmadyn-repo} to gather firmware images from the vendors listed in
Table~\ref{tab:vendor_overview}. We create a separate parser for each vendor 
website using XPath to parse a given root webpage. If the webpage contains 
an element matching a link to a firmware image, the parser will download the image.
Concurrently, the parser will record elements about the 
\emph{product name}, \emph{firmware version}, \emph{release time},
and \emph{device type}, when available. Other webpages referred by this 
webpage will then be recursively processed in a similar way.

While downloading firmware images, we only target files with an extension of 
\code{img}, \code{bin}, \code{rar}, \code{pkg}, \code{chk}, \code{tar}, \code{zip}, \code{stk}, and \code{rmt}. Setting rules on filename extension 
allows us to drop obviously non-firmware content like text scripts, PDF files, and Microsoft Office documents.
It also helps reduce the storage space needed to keep 
the downloaded data and their unpacked versions. 
To operate  within legal and ethical boundaries, we follow the procedure 
presented in~\cite{costin2014large}. We only download firmware images
released to the general public, and we obey the \code{robots.txt} directives
when presented. We will release all the crawlers upon publication of this paper.


\begin{figure}[!t]
        \centering
        \includegraphics[scale=0.18]{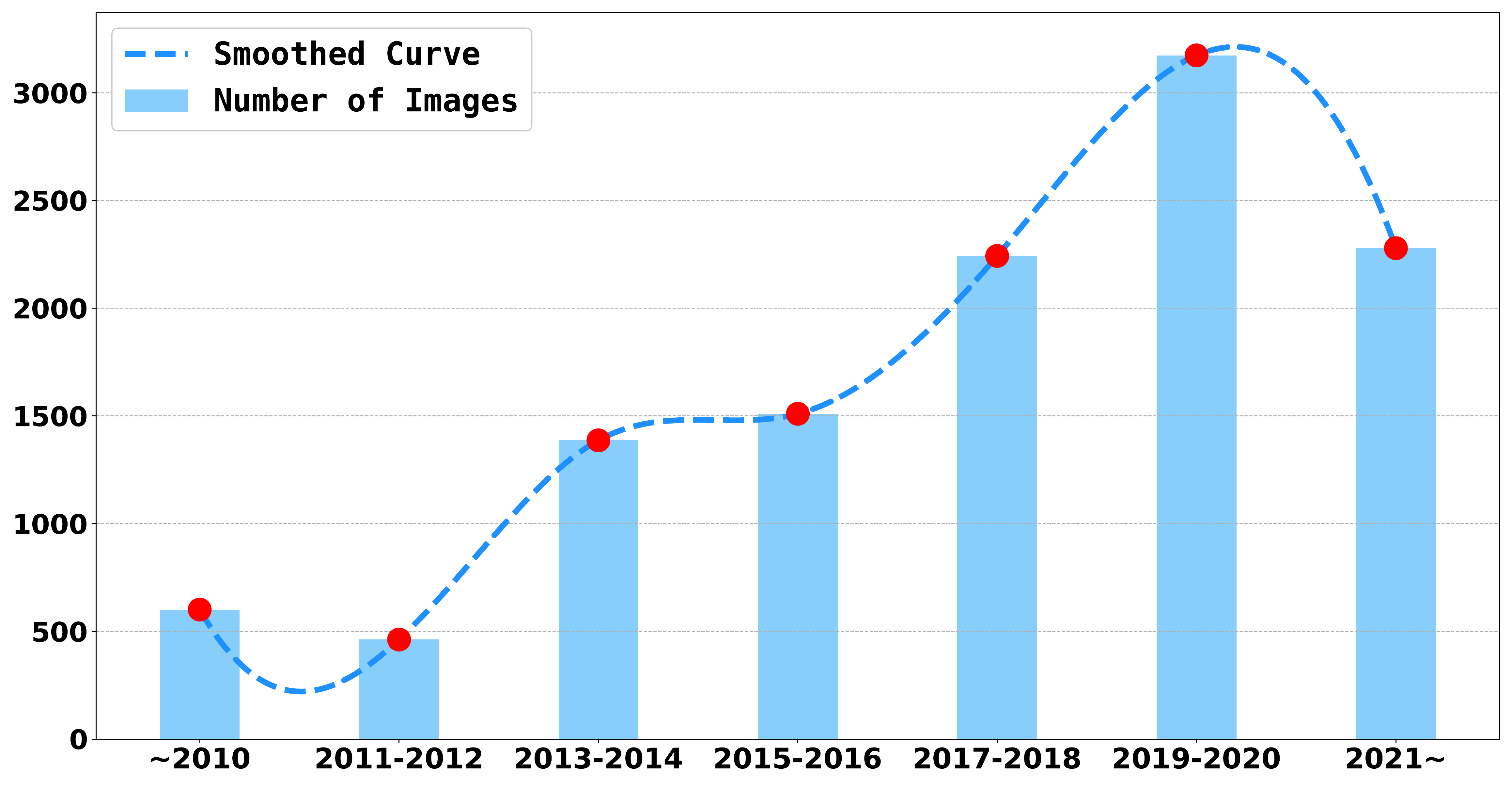}
        \vspace{-0.5em}
        \caption{Distribution of firmware images across release time. All images released before 2010 are aggregated into \code{$\sim$2010}.}
        \label{figure:dist-over-time}
        \vspace{-1em}
\end{figure}

\para{Results:} In total, we collected over 18,000 firmware images from 38 
vendors, as summarized in Table~\ref{tab:vendor_overview}. 
The release time of the firmware images spans two decades.
The earliest image was released back in 1998 (DES-3216 by D-Link),
and the most recent image just came out in 2021. 
Fig.~\ref{figure:dist-over-time} shows the distribution 
of the firmware images over their release time. Clearly,
more images were released in recent periods. 
The firmware images run on 4,000+ different products, 
covering many common types of device 
used in our daily life, as listed in Table~\ref{tab:fw-category} 
in the Appendix. Among the products, over 2,000 have multiple versions of firmware
available. This is important since it helps us with building an understanding of 
evolution over time. Overall, we envision this dataset has reasonable
scale, diversity, and representativeness to support our study.


\setlength\tabcolsep{3.5pt}
\begin{table}[t]
  \centering
  \caption{Statistical results of firmware images collected from popular embedded device vendors. \code{PMV} means product with multiple versions of firmware available. \code{Time Range} indicates the period where the images were released (\code{xx} means the year of \code{20xx} unless otherwise noted).}
  \vspace{-0.5em}
  \label{tab:vendor_overview}
  \footnotesize
  \begin{tabular}{l c c c c}
    \toprule
    \textbf{Vendor} & \textbf{\# of Images} & \textbf{\# of Products} & \textbf{\# of PMVs} & \textbf{Time Range} \\ \toprule
    Cerowrt    &  2        &   2        & 0   &   14 $\to$ 14   \\  \hline
    Haxorware  &  2        &   1         & 1   &  -   \\  \hline
    AT\&T          &  4        &   3      & 1    &  -     \\  \hline
    360        &  5        &   1      & 1    &   17 $\to$ 17   \\  \hline
    Actiontec      &  6        &   5     & 1     &  -    \\  \hline
    Buffalo    &  6        &   3       & 2   &  16 $\to$ 18    \\  \hline
    camius     &  6        &   6        & 0  &  -    \\  \hline
    GOCloud    &  8        &   5         & 2   &  19 $\to$ 21     \\  \hline
    Phicomm        &  13       &   7       & 4   &  16 $\to$ 18   \\  \hline
    ZyXEL      &  15       &  7        & 4    &  17 $\to$ 21   \\ \hline
    CenturyLink    &  18       &   18  & 0    &  -   \\  \hline
    Polycom        &  21       &   4   & 3  &  18 $\to$ 19    \\  \hline
    u-blox      &  31       &  25       & 6    &  16 $\to$ 21    \\  \hline
    TENVIS     &  41       &  16       & 2     &    12 $\to$ 14  \\  \hline
    MikroTik   &  49       &  16       & 13     &   -  \\  \hline
    Foscam         &  83       &   36    & 22    &  13 $\to$ 18    \\  \hline
    AVM        &  107      &   54      & 44    &    -  \\  \hline
    RouterTech &  144      &  15         & 10     &  06 $\to$ 11    \\  \hline
    Belkin     &  165      &  109       & 33    &   -   \\  \hline
    Linksys    &  166      &  132      & 38   &  01 $\to$ 21     \\  \hline
    Mercury    &  169      &  142      & 35   & 09 $\to$ 20    \\  \hline
    Supermicro &  187      &  186      & 1       &      - \\  \hline
    Digi           &  214      &   100         & 56   & -       \\  \hline
    NETCore    &  255      &  229      & 21     &  20 $\to$ 21   \\  \hline
    Moxa           &  400      &   315      & 53   &   04 $\to$ 18    \\  \hline 
    TRENDnet   &  409      &  365      & 46   &  12 $\to$ 21    \\  \hline
    Tenda          &  467      &   179         & 110   &  09 $\to$ 18      \\ \hline
    Ubiquiti       &  512      &   206      & 165    &    07 $\to$ 18    \\  \hline
    QNAP           &  576      &   27       & 22    & 16 $\to$ 18     \\  \hline
    Hikvision   &  607      &  112         & 77   &  14 $\to$ 21   \\  \hline
    Synology       &  672      &   117      &99      &  -     \\  \hline
    
    TomatoShibby   &  692      &   7          & 7  &    14 $\to$ 16    \\  \hline
    Tp-Link-zh  &  992      &  566      & 187    &  08 $\to$ 23    \\  \hline
    ASUS           &  1099     &   179      & 146       &  06 $\to$ 18   \\  \hline
    D-Link      &  1172     &  260        & 218  &   98$^1$ $\to$ 20   \\  \hline
    Tp-Link-en  &  1186     &  274      & 225    &   -   \\  \hline
    NETGEAR    &  3682     &  663      & 449     &   - \\  \hline
    OpenWrt    &  3837     &  73       & 70     &    20 $\to$ 21    \\  \bottomrule
    \textbf{Total}      &  \textbf{18,020}    &  \textbf{4,470}     & \textbf{2,174}    &  - \\  
    \bottomrule
    \multicolumn{5}{ l }{$^{1}$~``98'' here means 1998.
    }
  \end{tabular}
  \vspace{-1em}
\end{table}
\setlength\tabcolsep{6pt}


\setlength\tabcolsep{4pt}
\begin{table*}[t!]
  \centering
  \caption{Unpacking results. The column of \code{.config}
  shows the number of kernels with \code{.config} file identified. 
  The column of \code{converted} shows the number of kernels that 
  can be converted to ELF. Vendors with no image unpacked are highlighted.}
  \vspace{-0.5em}
  \label{tab:extracted-result}
  \footnotesize
  \begin{tabular}{ l | c| c c c c c c c c| c c| c c c}
    \toprule
    \multirow{2}{*}{\textbf{Vendor}} & \multirow{2}{*}{\textbf{\# of Images}} & \multicolumn{8}{c|}{\textbf{Unpacked Images}}& \multicolumn{2}{c|}{\textbf{Filesystems}} & \multicolumn{3}{c}{\textbf{Linux Kernels}}\\ \cline{3-15}
    & & \code{Total} & \code{ARM} & \code{AArch64} & \code{MIPS} & \code{x86} & \code{x64} & \code{PowerPC} & \code{Other} & \code{Total} & ELF (\code{k}) & \code{Total}  & \code{.config} & \code{converted} \\  \toprule
    Cerowrt & 2 & 2 & 0 & 0 & 2 & 0 & 0 & 0 & 0 & 2 & 0.4 & 0 & 0 & 0 \\\hline
    Haxorware & 2 & 1 & 1 & 0 & 0 & 0 & 0 & 0 & 0 & 1 & 0.2 & 0 & 0 & 0\\\hline
    AT\&T & 4 & 4 & 0 & 0 & 4 & 0 & 0 & 0 & 0 & 4 & 0.6 & 0 & 0 & 0 \\ \hline
    360 & 5 & 4 & 0 & 0 & 4 & 0 & 0 & 0 & 0 & 4 & 0.5 & 4 & 0 & 2 \\ \hline
    Actiontec & 6 & 5 & 2 & 0 & 3 & 0 & 0 & 0 & 0 & 5 & 0.4 & 0 & 0 & 0 \\ \hline
    Buffalo & 6 & 4 & 0 & 0 & 4 & 0 & 0 & 0 & 0 & 4 & 0.5 & 4 & 0 & 2 \\\hline
    Camius & 6 & 6 & 0 & 0 & 6 & 0 & 0 & 0 & 0 & 6 & 0.5 & 6 & 0 & 6 \\\hline
    GOCloud & 8 & 7 & 0 & 0 & 0 & 7 & 0 & 0 & 0& 7 & 0.9 & 0 & 0 & 0\\\hline
    Phicomm & 13 & 8 & 2 & 0 & 6 & 0 & 0 & 0 & 0 & 8 & 1.9 & 3 & 1 & 3 \\ \hline
    ZyXEL & 15 & 8 & 8 & 0 & 0 & 0 & 0 & 0 & 0 & 7 & 0.8 & 7 & 0 & 3 \\\hline
    CenturyLink & 18 & 7 & 0 & 0 & 7 & 0 & 0 & 0 & 0 & 7 & 0.8 & 2 & 0 & 1 \\ \hline
    Polycom & 21 & 16 & 0 & 0 & 0 & 0 & 0 & 0 & 16 & 0 & 0 & 16 & 16 & 0 \\ \hline
    \rowcolor{lightgray}
    u-blox & 31 & 0 & 0 & 0 & 0 & 0 & 0 & 0 & 0 & 0 & 0 & 0 & 0 & 0\\\hline
    TENVIS & 41 & 31 & 27 & 0 & 4 & 0 & 0 & 0 & 0 & 25 & 0.9 & 31 & 0 & 0 \\\hline
    MikroTik & 49 & 32 & 8 & 0 & 12 & 4 & 0 & 4 & 4 & 32 & 4.3 & 0 & 0 & 0\\\hline
    Foscam & 83 & 10 & 10 & 0 & 0 & 0 & 0 & 0 & 0 & 0 & 0 & 10 & 0 & 0 \\ \hline
    AVM & 107 & 22 & 15 & 0 & 7 & 0 & 0 & 0 & 0 & 22 & 5.0 & 0  & 0 & 0\\ \hline
    RouterTech & 144 & 143 & 0 & 0 & 143 & 0 & 0 & 0 & 0 & 143 & 25.8 & 142 & 0 & 0 \\\hline
    Belkin & 165 & 67 & 6 & 1 & 60 & 0 & 0 & 0 & 0 & 60 & 7.9 & 60 & 0 & 33 \\ \hline
    Linksys & 166 & 115 & 58 & 0 & 57 & 0 & 0 & 0 &  0 & 74 & 17.1 & 101 & 24 & 75 \\\hline
    Mercury & 169 & 27 & 0 & 0 & 27 & 0 & 0 & 0 & 0 & 27 & 1.5 & 27 & 0 & 27\\\hline
    Supermicro & 187 & 187 & 185 & 0& 0 & 0 & 0 & 0 & 2 & 5 & 1.3 & 187 & 7 & 9\\\hline
    Digi & 214 & 3 & 0 & 0 & 3 & 0 & 0 & 0 & 0 & 3 & 1.5 & 5 & 1 & 2 \\ \hline
    NETCore & 255 & 153 & 1 & 0 & 152 & 0 & 0 & 0 & 0 & 152 & 10.2 & 138 & 1 & 85 \\\hline
    Moxa & 400 & 107 & 83 & 0 & 20 & 0 & 0 & 0 & 4 & 107 & 32.0 & 0 & 0 & 0 \\ \hline
    TRENDnet & 409 & 169 & 38 & 0 & 130 & 0 & 0 & 0 & 1 & 142  & 15.3 & 158 & 3 & 70 \\\hline
    Tenda & 467 & 252 & 64 & 0 & 188 & 0 & 0 & 0 & 0 & 252 & 33.6 & 142 & 0 & 118 \\ \hline
    Ubiquiti & 512 & 479 & 137 & 0 & 342 & 0 & 0 & 0 & 0 & 479 & 204.7 & 449 & 59 & 436 \\ \hline
    QNAP & 576 & 297 & 0 & 0 & 0 & 297 & 0 & 0 & 0 & 297 & 296 & 0 & 0 & 0\\ \hline
    Hikvision & 607 & 190 & 189 & 1 & 0 & 0 & 0 & 0 & 0 & 0 & 0 & 190 & 41 & 186 \\\hline
    Synology & 672 & 671 & 346 & 24 & 0 & 15 & 250 & 36 & 0 & 671 & 1375.4 & 0 & 0 & 0\\ \hline
    TomatoShibby & 692 & 692 & 118 & 0 & 574 & 0 & 0 & 0 & 0 & 692 & 127.8 & 314 & 0 & 23 \\ \hline
    Tp-Link-zh & 992 & 494 & 175 & 0 & 319 & 0 & 0 & 0 & 0 & 464 & 65.7 & 385 & 53 & 325 \\\hline
    ASUS & 1,099 & 1,069 & 388 & 0 & 678 & 0 & 0 & 0 & 3 & 1,069 & 273.2 & 438 & 54 & 288 \\ \hline
    D-Link & 1,172 & 134 & 52 & 0 & 68 & 0 & 0 & 14 & 0 & 86 & 15.9 & 116 & 11 & 92\\\hline
    Tp-Link-en & 1,186 & 660 & 107 & 3 & 548 & 0 & 0 & 2 & 0 & 654 & 76.3 & 565 & 43 & 544 \\\hline
    NETGEAR & 3,682 & 1,474 & 443 & 12 & 932 & 5 & 31 & 51 & 0 & 980 & 173.9 & 1,293 & 269 & 957\\\hline
    OpenWrt & 3,837  & 3,335 & 276 & 18 & 3,021 & 0 & 0 & 20 & 0 & 2,546 & 191.2 & 3,184 & 0 & 0\\\bottomrule
    \textbf{Total} & \textbf{18,020} & \textbf{10,685} & \textbf{2,540} & \textbf{58}  & \textbf{7,321}  & \textbf{328} & \textbf{281} & \textbf{127} & \textbf{27} & \textbf{9,037} & \textbf{2,964} & \textbf{7,977} & \textbf{581} & \textbf{3,287} \\\bottomrule
  \end{tabular}
  \vspace{-1em}
\end{table*}
\setlength\tabcolsep{6pt}

\subsection{Unpacking Firmware Images}
\label{subsec:unpack}

The next step focuses on unpacking each firmware and 
extracting the binary programs and Linux kernel inside. 
Linux-based firmware is a concatenated archive of different
parts of a Linux system. As depicted in Fig.~\ref{tab:fw_img} 
in the Appendix, the archive usually includes one or more
filesystem partitions, a Linux kernel, a bootloader, and 
various configurations and other data files. 
Given a firmware image, which is often compressed, 
we first decompress it according to the compression 
algorithms (\code{zip}, \code{bzip2}, \code{gzip}, 
\code{tar}, \code{rar}, \etc). Customized compression 
algorithms are not handled due to a lack of specifications.
We then unarchive the decompressed image to extract 
filesystems and the Linux kernel.

\para{Extracting Filesystems:} We reuse \firmadyne~\cite{chen2016towards}, 
a tool built upon \binwalk~\cite{binwalk}, to extract filesystems from 
firmware images. Besides using manually-created signatures to 
locate complete filesystems, \firmadyne also searches for standard 
directories under the root directory (\eg\xspace{}\code{bin}, \code{sbin}, 
\code{lib}, \etc). The filesystems and directories are then
recursively traversed to identify binary programs/libraries
in ELF format.

\para{Extracting Linux Kernels:} When unpacking a firmware image, \firmadyne
can identify Linux kernels based on signatures inherited from \binwalk. 
However, the signatures are too specific, filtering out many
kernels customized for embedded devices. We extended 
the signatures based on patterns observed in our dataset.
Doing so enables us to identify 58.3\% more Linux kernels. 

The kernels extracted by \firmadyne are in raw data format, which 
cannot be further parsed and analyzed. To address this issue,
we convert the kernels to fully analyzable ELF 
files with the help of \vmlinuxtool~\cite{vmlinux-elf}. 
At the high level, \vmlinuxtool identifies symbol tables (\code{kallsyms})
in a given kernel to identify functions and then reorganizes them 
into an ELF file. More technical details about \vmlinuxtool
can be found in its manual~\cite{vmlinux-elf}.

To identify mitigations in a Linux kernel, we often need the 
configuration file used to build the kernel (commonly
known as \code{.config}). To find the \code{.config} file, 
we run \binwalk to recursively extract files from the raw kernel
and check whether they are \code{.config} using the \code{file}
utility.

\para{Results:} We consider a firmware image unpacked if we extract any ELF files 
or a Linux kernel from the image. Based on this criterion, 10,685 out of 18,020 
firmware images are unpacked, as reported in Table~\ref{tab:extracted-result}. 
The success rate is 59.3\%, which is comparable to previous research 
(26,275 out of 32,256 images~\cite{costin2014large} and 8,893 out of 23,035 images~\cite{chen2016towards}). The unpacked images are from 37 vendors 
and span all the major architectures (ARM, AArch64, MIPS, x86, x64, and PowerPC).
For the 7,335 images that we cannot unpack, 4,277 are either non-Linux based 
or encrypted, which are out of our consideration. The other 3,058 contain
nothing that \firmadyne can recognize.


From the 10,685 unpacked images, we collected 9,037 filesystems 
(spanning 34 vendors). The filesystems contain around 3,000k ELF binaries.
More details about the binaries are presented in \S\ref{sec:user-space}. 
From the unpacked images, we also extracted 7,977 Linux kernels in 
the format of raw data. The kernels include 99 distinct versions, ranging 
from \code{v2.0.40} to \code{v4.14.221}. \vmlinuxtool converted 3,287 
of them to ELF files with symbols. It failed to convert the remaining kernels
because \code{kallsyms} is not available (4,672 kernels) or 
the architectures are unknown (18 kernels). The identification 
of \code{.config} files is less rewarding. We only extracted \code{.config} 
files from 581 kernels. This is 
understandable since \code{.config} is typically not needed 
for deployment.




\section{Identification of Mitigations}
\label{sec:mitigation}


\subsection{Identifying Mitigations in User-space Binaries}
\label{subsec:binary-mitigation}

Since the 2000s, the security of binary software has made 
leaps forward through the adoption of a variety of mitigations
at the compiler and OS levels. Table~\ref{tab:user-space-defense} 
highlights the mitigations we aim to identify. We included all the mitigations that are both integrated into standard compiling/building toolchains of Linux systems and found active on modern Linux distribution. The mitigations also represent the ones concerned by the hacking communities. For instance, Pwntools~\cite{pwntools}, a popular exploit framework, considers the same set of mitigations. In the following, we introduce 
each of them and explain how we detect their presence.


\subsubsection{Stack Canaries}
\label{subsec:canary}
Stack Canaries, also known as stack guards~\cite{cowan1998stackguard}, 
are used to provide defense against 
stack overflows. This mitigation is implemented by compilers (\eg GCC) 
via inserting special code at the entry and exit points of functions. 
At function entry, a secret, random canary value is saved at the top of the stack 
separating the return address from the stack frame.
At function exit, the canary value is checked. If the canary value is not changed, 
the function returns to the caller. Otherwise, the function calls 
\code{\_\_stack\_chk\_fail}, a routine provided by the C library, 
to terminate the execution and report an error.

Existing tools, including \pwntools~\cite{pwntools} and 
\checksec~\cite{checksec}, provide modules to detect Stack Canaries.
They report the deployment of Stack Canaries when the symbols or 
relocation entries (\ie Global Offset Table, or GOT, entries) 
contain \code{\_\_stack\_chk\_fail}. This approach works well on 
binaries that are dynamically linked or non-stripped. However, 
many binaries running on embedded devices are statically linked 
and stripped. Accordingly to our analysis, existing  tools 
completely failed to detect Stack Canaries deployed in hundreds
of those binaries in our dataset.

To detect Stack Canaries in statically-linked, stripped binaries,
we use a generic heuristic. We observe that, when a 
stack violation is detected, \code{\_\_stack\_chk\_fail} 
prints a error message starting with ``\code{*** stack smashing 
detected ***}''. The message is not influenced by optimization level, 
CPU architecture, or the type of binary. We, thus, search for 
that string in a given binary. Once found, we disassemble the binary to 
locate the function (\ie\xspace{}\code{\_\_stack\_chk\_fail}) 
that uses the string. If the function is identified and called
by other functions, we consider Stack Canaries are deployed. 
Listing~\ref{lst:static_canary} in the Appendix shows an 
example of \code{\_\_stack\_chk\_fail} in 
a statically-linked ARM32 binary, which shows a pointer
to the above string is used.

\subsubsection{Relocation Read-Only}
Relocation Read-Only (RELRO) is a defense measurement against GOT 
hijacking~\cite{spbypass:phrack00}, applied when linking binaries. 
GOT holds the addresses of variables and functions that are unknown 
during linking but relocated at run-time (\eg  variables and functions imported
from libraries). In contemporary systems, the GOT often splits into two sections: 
\code{.got} and \code{.got.plt}, with the latter being used by the
Procedure Linkage Table (PLT). Briefly, the PLT
includes code that enables \textit{lazy binding} of external functions. 
Specifically, when an external function is called for the first time, 
code in the PLT executes and calls the linker for resolving the symbol 
and writing the function's address to the GOT entry (hence, \textit{lazy}). 
Next time the external function is called, the PLT code directly jumps to 
the address in the GOT entry. Because symbols are resolved at
run time, the GOT needs to be writable, making it vulnerable 
to attacks that corrupt the GOT entries to hijack program execution. 
 



RELRO~\cite{RELRO} aims to turn parts or all of the GOT 
read only (RO) to protect the 
function pointers from overwrites. This is done by resolving 
symbols within the GOT at load time and remapping it to 
RO before the program executes. There are two versions of RELRO: \textit{partial} 
and \textit{full} RELRO. Partial RELRO only protects the \code{.got} section,
which stores offsets to symbols of variables, leaving \code{.got.plt} 
writable to perform lazy binding. 
Full RELRO keeps a single \code{.got} section and protects all of it, 
requiring that all symbols are resolved in the beginning. 
Fig.~\ref{fig:relro} in the Appendix shows the difference between no
RELRO, partial RELRO, and full RELRO.



RELRO is implemented by the linker based on the metadata found in ELF binaries,
which specifies the GOT to be mapped with the designated
protections (\eg RO). We observe that for RELRO to be present, 
\code{.got} needs to be mapped to a RO segment (\code{GNU\_RELRO}), 
while the concurrent presence of a \code{.got.plt} in a writable 
segment indicates that we only have partial RELRO. Additionally, 
binaries with full RELRO require external symbols to be resolved 
at load time, which is enabled by one of the following linking 
flags being present in the \code{.dynamic} section: 
\code{BIND\_NOW, DT\_BIND\_NOW, DF\_1\_NOW}. Our study relies on the appearance of
\code{.got} and {.got.plt}, the protections of their segments, 
and the linking flags, as specified above, to identify RELRO.

\subsubsection{Non-executable Stack}
Most modern processors~\cite{abbasi2019challenges}, including 
microprocessors of the ARM-Cortex-R and ARM-Cortex-M families, 
several MIPS32, and most PowerPC processors, support data-execution 
prevention (DEP)~\cite{dep}. DEP is a feature that prevents the 
execution of instructions from protected memory segments, 
in particular, segments that are also writable. 


The use of DEP for the program stack is most crucial to 
defang stack overflow vulnerabilities, commonly 
known as Non-executable (NX) Stack~\cite{Non-executableStack}.
To enable NX Stack, 
binaries need to specify that the stack is NX explicitly. 
To safely detect the adoption of NX stack, 
we need to confirm DEP support from the hardware and
the presence of \code{PT\_GNU\_STACK} in the program header
for the stack segment (which mandates the stack is NX). However,
due to the lack of hardware specifications, 
our study only checks the program header. This should 
not cause many problems since DEP is a pretty standard feature
on processors that can run Linux-based systems.


\subsubsection{Fortify Source}
The standard C library (libc) includes many unsafe functions 
that can lead to overflows when misused. Fortify Source
~\cite{FortifySource} is a 
defense measurement activated by compilers like GCC to check on known 
unsafe functions in libc. These mainly include functions that copy or
write data to a destination buffer without limiting 
the number of bytes (\eg\xspace{}\code{strcpy}, \code{strcat}, \code{memcpy}, \etc). 
Fortify Source replaces those functions with safer versions
that perform size checks. Fig.~\ref{fig:fortify} in the Appendix 
depicts one such example, where two functions are replaced with 
their safer counterparts.



Existing tools detect Fortify Source based on symbols
or relocation entries of replacement functions in the form of 
\code{\_\_*\_chk}. Similar to the identification of Stack 
Canaries, this approach is effective with dynamically linked 
or non-stripped binaries because \code{\_\_*\_chk} are defined in 
libc and their symbols are imported. 
However, it cannot handle statically-linked, stripped binaries.
To this end, we again apply the heuristic we used to detect
Stack Canaries. The replacement functions output a constant 
message ``\code{*** buffer overflow detected ***}'', when 
detecting violations. We follow this message as an indicator 
to locate \code{\_\_*\_chk} functions and, in turn, identify 
the adoption of Fortify Source.


\subsubsection{Position-Independent (or ASLR-Capable) Code}
Address Space Layout Randomization (ASLR)~\cite{ASLR} is a seminal 
defense for mitigating exploitation. ASLR mandates that each time 
a program executes, the code segment, the stack, the heap, and 
the libraries are located at a randomly selected 
offset in memory. 
Besides OS support, ASLR requires binaries (programs or libraries) 
to be compiled as Position Independent Executable (PIE) and Position 
Independent Code (PIC), or otherwise, \textit{relocatable} code. 
Whether a binary is position-independent or relocatable is 
indicated by its program header. Specifically, position-independent or
relocatable binaries have type \code{ET\_DYN}. Otherwise, 
they will have type \code{EX\_EXEC}. This enables us to 
identify ASLR-capable binaries by checking their program headers.

\setlength\tabcolsep{4.6pt}
\begin{table}[!t]
  \centering
  \caption{Adoption rates of user-space mitigations (\%). The best result for
  each mitigation is highlighted. \textbf{Ave (Vendor)} shows results 
  averaged on vendors while \textbf{Ave (Binary)} indicates results 
  calculated on all binaries.}
  \vspace{-0.5em}
  \label{tab:mitigation-result}
  \footnotesize
  \begin{tabular}{ l| c| c| c| c| c| c}
 \toprule
 \textbf{Vendors} & \textbf{ELF (k)} & \textbf{Canary} & \textbf{RELRO}  & \textbf{NX} & \textbf{Fortify} & \textbf{PIE} \\\toprule
Haxorware& 0.2     & 0     & 0     & 0    & 0     & 14.9  \\ \hline
Actiontec & 0.4    & 0.5    & 0      & 47.2     & 0.5        & 13.4    \\ \hline
Cerowrt   & 0.4       & 0         & 0     & 0        & 0       & 9.8    \\ \hline
360    & 0.5       & 60.0      & 0     & 0        & 0       & 8.9     \\ \hline
Buffalo   & 0.5       & 0         & 0      & 45.8     & 0       & 6.0    \\ \hline
Camius    & 0.5       & 11.9    &  0  & 92.1     & 1.3        & 11.9     \\ \hline
AT\&T  & 0.6       & 0         & 0      & 0     & 0       & 6.3     \\ \hline
CenturyLink  & 0.8       & 0         & 0      & 0      & 0       & 0.6     \\ \hline
Zyxel  & 0.8       & 1.0         & 0      & 97.3     & 0.9        & 11.6    \\ \hline
GOCloud   & 0.9       & 0         & 0      & 98.2    & 0       & 14.9    \\ \hline
TENVIS    & 0.9       & 0         & 0      & 0        & 0       & 34   \\ \hline
Supermirco   & 1.3       & 19.4      & 3.2    & 97.8    & 16.1       & 18.5    \\ \hline
Digi   & 1.5       & 0         & 0     & 3.5      & 0       & 18.5     \\ \hline
Mercury   & 1.5       & 0         & 0      & 0        & 0       & 31.5   \\ \hline
Phicomm   & 1.9       & 0.1       & 0.8      & 21.2       & 0       & 47.2    \\ \hline
MikroTik  & 4.3       & 0.2       & 7.9      & 81.0       & 0.07        & 5.8   \\ \hline
AVM    & 5.0      & \cellcolor{lightgray}{81.5}      & \cellcolor{lightgray}{89.4}  & 95.6     & 0.04       & \cellcolor{lightgray}{90.8}    \\ \hline
Belkin    & 7.9       & 0.2       & 3.8      & 7.4      & 1.6        & 11.0   \\ \hline
NETCore   & 10.2      & 11.3      & 0.02    & 0.06     & 0.2        & 16.4     \\ \hline
TRENDnet  & 15.3      & 0.4       & 0.3     & 10.1     & 0.05       & 13.6     \\ \hline
Dlink  & 15.9      & 0.4       & 0.4     & 30.4     & 0.04       & 9.1     \\ \hline
Linksys   & 17.1      & 0.5       & 3        & 60.4     & 0.8        & 9.0    \\ \hline
RouterTech   & 25.8      & 0         & 0      & 0       & 0       & 15.0     \\ \hline
Moxa   & 32.0        & 39.3      & 15.0    & 75.7    & 35.5       & 31.8 \\ \hline
Tenda  & 33.6      & 0.6       & 2.3  & 30.5     & 0.01       & 11.7     \\ \hline
Tp-Link-zh   & 65.7      & 2.9       & 0.4     & 38.7    & 0.1        & 18.3    \\ \hline
Tp-Link-en   & 76.3      & 0.5       & 0.9    & 36.6     & 0.6        & 21.5     \\ \hline
TomatoShibby & 127.8     & 0.1       & 1.0       & 23.2     & 0       & 8.4     \\ \hline
NETGEAR   & 173.9     & 2.2       & 4.4     & 55.9     & 0.5        & 11.4     \\ \hline
OpenWrt   & 191.2     & 0         & 0    & \cellcolor{lightgray}{99.9}     & 0       & 0   \\ \hline
Ubiquiti  & 204.7     & 6.7       & 1.0   & 15.6     & 25.0         & 9.5    \\ \hline
ASUS   & 273.2     & 1.3       & 1.4    & 46.8     & 0.05       & 8.3  \\ \hline
QNAP   & 296.0       & 80.1      & 3.1      & 99.2     & 1.4        & 7.7   \\ \hline
Synology  & 1375.4       & 43.6      & {36.7}     & 99.5     & \cellcolor{lightgray}{43.5}       & 13.5   \\ \bottomrule

 \textbf{Ave (Vendor)} & 87.2  & 10.7     & 5.2  & 41.5   & 3.5      & 16.5   \\  \hline
 \textbf{Ave (Binary)} & -  & 29.7     & 18.3  &76.2    & 22.5      & 11.6  \\  \bottomrule
 \textbf{Debian} & 34.0  &  85.3    & 98.1  &  99.7  &    55.6   &    94.0 \\  \bottomrule
  \end{tabular}
  \vspace{-1em}
\end{table}

\setlength\tabcolsep{6pt}

\subsection{Identifying Mitigations in the Kernel}
\label{subsec:kernel}

Linux kernel has been gradually incorporating various attack
mitigations since version 2.6. We examined all popular kernel mitigations~\cite{kernel-defense} and targeted those (i) applicable to deployed systems (ii) active in modern Linux distributions and (iii) released over three years ago (such that the vendors have sufficient time to deploy them). Table~\ref{tab:kernel-space-defense} summarizes the ones we finally picked. Identification of these
kernel-level mitigations can be done more systematically.
As per Table~\ref{tab:kernel-space-defense}, the presence of the mitigations
can be identified based on the architecture, the kernel version, 
and the building configurations. Since \vmlinuxtool already 
provides the architecture information, we do not have to worry about it.
In the following, we describe the recovery of kernel version 
and the identification of mitigations with and without 
the configuration files. 

\subsubsection{Recovery of Kernel Version}
\label{subsec:kernel-version}
When \code{.config} is recovered, the kernel version is explicitly 
documented within. However, as we pointed out before, the \code{.config} 
file is not always available.  When the \code{.config} is missing, 
we instead search for
string constants within the kernel image to infer its version. 
For instance, kernels frequently include string resembling
\textit{Linux version 2.6.36 (root@automake) 
(gcc version 4.6.3 (Buildroot 2012.11.1) ) \#2 Fri Jan 20 15:50:29 CST 2017}, 
which gives explicit information about the kernel version.


\subsubsection{Analysis with Building Configuration}
\label{subsec:config}
An option in the \code{.config} file being selected 
(\eg\xspace{}\code{CONFIG}\code{\_HAVE}\code{\_CC}\code{\_STACK\_PROTECTOR=y}) 
means the corresponding
feature is enabled. In contrast, an un-selected option, 
indicated by its appearance in a line starting with ``\#'' 
or its absence in the file, means the feature is not enabled. 
With the support of the \code{.config} file, 
we can easily determine whether a target mitigation is
activated in the kernel by checking the associate options
specified in Table~\ref{tab:kernel-space-defense}.  

\subsubsection{Analysis without Building Configuration}
\label{subsec:elfkernel}

When the \code{.config} file is missing, we may still measure
the presence of many mitigations based on the ELF file converted 
from the kernel. Consider Stack Protector as an example.
We can detect its presence by the existence and usage of 
indicator function \code{\_\_stack\_chk\_fail}. 
Similarly, Fortify Source, Vmap Kernel Stack, USERCOPY, Heap Freelist Obfuscation, 
Executable Memory Protection, and KASLR can be respectively 
detected using \code{**\_chk}, \code{free\_vm\_stack\_cache}, \code{usercopy\_warn},
\code{freelist\_state\_initialize}, \code{mark\_rodata\_ro},
and \code{rotate\_xor} as indicator functions. Finding 
indicator functions in the converted ELF is straightforward since 
\vmlinuxtool already recovered the symbols. 

\section{Measuring Adoption of User-Space Mitigations}
\label{sec:user-space}

We run the identification of user-space mitigations on the 
3,000k binaries described in \S\ref{sec:data}. Binaries 
released before a mitigation are excluded from the analysis
of that mitigation. 
For NX Stack and PIE, only executables are considered
because the two mitigations are less meaningful for libraries. 
For other mitigations, all binaries, including both
executables and libraries, are considered. In addition, 
full RELRO and partial RELRO are aggregated together. 
To build a baseline for references, we further 
run the identification on 34k ELF binaries extracted from 
7,483 Debian packages located in the stable distribution
for desktop. The Debian binaries mostly run on x86/x64 architectures.
The measurement results are summarized in Table~\ref{tab:mitigation-result}. 

\setlength\tabcolsep{3.5pt}
\begin{table}[!t]
  \centering
  \caption{Adoption rates of user-space mitigations by different types of binary (\%). \code{Exe} and \code{Lib} stand for executable and library, respectively; ``-'' indicates the mitigation is not applicable or not meaningful.} 
  \vspace{-0.5em}
  \label{tab:elftype-mitigation}
  \footnotesize
  \begin{tabular}{ l | c| c| c| c| c| c| c}
 \toprule
 & \textbf{Type} & \textbf{ELF (k)} & \textbf{Canary} & \textbf{RELRO} & \textbf{NX} & \textbf{Fortify} & \textbf{PIE} \\\toprule

\multirow{3}{*}{\rotatebox{90}{\code{Emd.}}} & Dynamic Exe &   1,340.3   &  30.9   &  15.4 &  75.5   &  26.1    & 11.7   \\ \cline{2-8}
& Dynamic Lib   &  1,615.0  &  28.8  &   20.7   &  -   &    19.6     &  -  \\ \cline{2-8}
& Static Exe  &    7.9    &   8.8      &  3.9    &    42.4    &    8.4   &   0  \\ \toprule
\multirow{3}{*}{\rotatebox{90}{\code{Debian}}} & Dynamic Exe   &  20.0 & 89.9   &  98.7   &  99.9    &  75.2      &   94.0 \\\cline{2-8}
& Dynamic Lib   &  14.0 & 79.8   &  98.1   &  -    &  30.2     &  - \\\cline{2-8}
& Static Exe   &  0.2 &  24.1  &   38.8    &    87.1 & 24.6    &  0 \\
\bottomrule

  \end{tabular}
  \vspace{-1em}
\end{table}

\setlength\tabcolsep{6pt}

\subsection{Analysis of Results}
Overall, the adoption rates of attack mitigations by embedded
binaries are not high, significantly falling behind the desktop
binaries (represented by Debian). Stack Canaries, one of the most 
common mitigations in desktop binaries, are only applied to 29.7\% 
of the embedded binaries. Zooming into the results, even this 29.7\% adoption 
rate is likely an overestimation of the general reality as it is 
exaggerated by the large number of binaries with Stack Canaries 
from QNAP and Synology. Without counting the two vendors, 
the adoption rate drops to 3.28\%. The situation of RELRO and PIE 
is similar. The adoption rates are  98.1\% and 
94.0\% on desktop binaries, dramatically dropping to 18.3\% 
and 11.6\% on embedded binaries. 

\begin{figure}[!t]
        \includegraphics[width=0.48\textwidth]{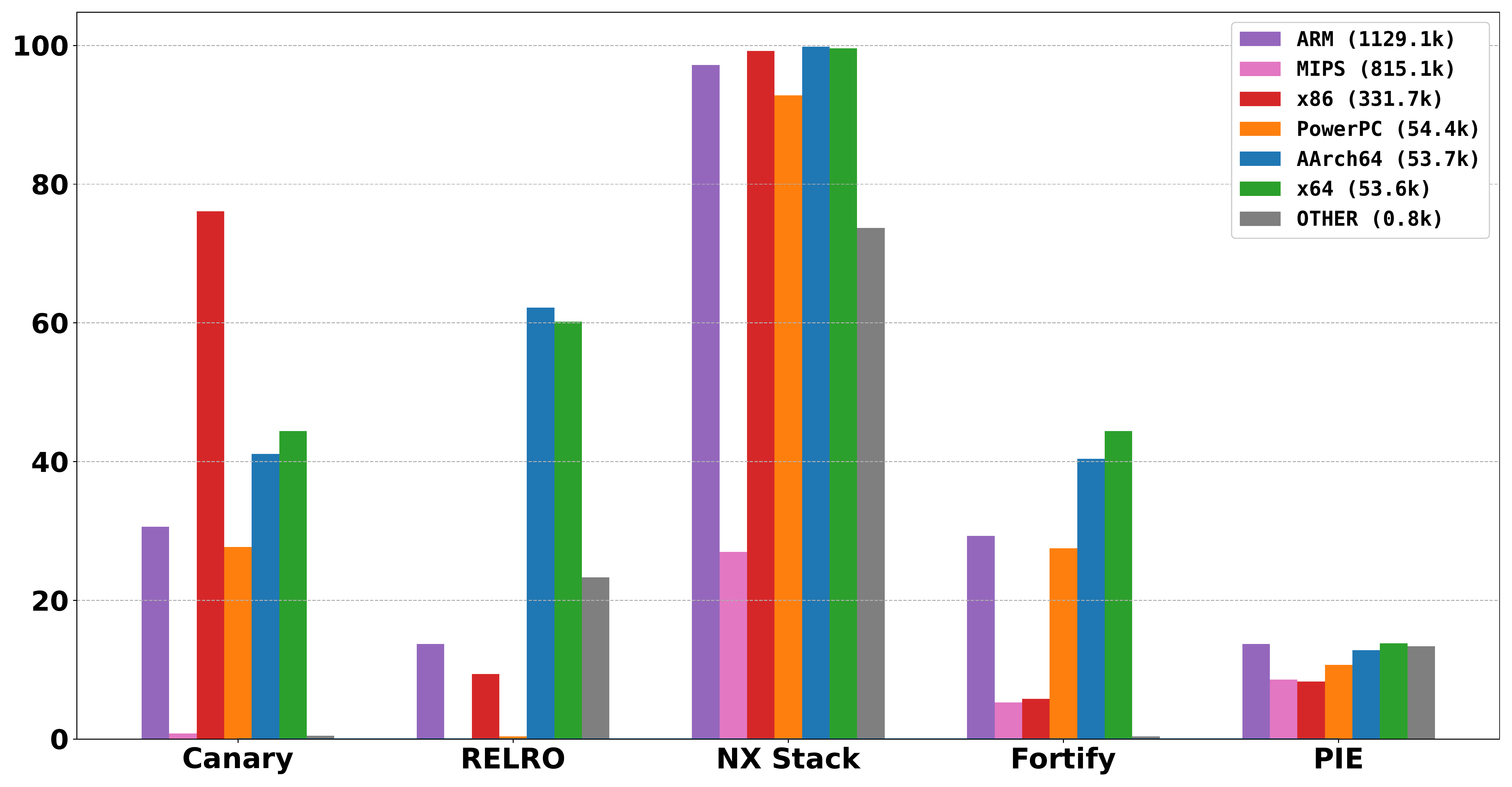}
        \vspace{-1em}
        \caption{Adoption rates of user-space mitigations by binaries running on different architectures. The numbers in the legend represent
        the number of binaries with each architecture.}
        \label{tab:mitigation-arch}
        \vspace{-1.5em}
\end{figure}

Regarding the adoption of Fortify Source, embedded binaries 
also fall behind desktop binaries but to a less significant
extent (22.5\% \vs 55.6\%). A factor contributing to
the gap is the broad use of uClibc by embedded binaries (about 22\%). 
Unlike Glibc, uClibc does not support Fortify Source. 
Without considering binaries that use uClibc, the adoption 
rate of Fortify Source raises to 36.5\%. 
The results of NX Stack are somewhat surprising. NX Stack,
a straightforward, no-cost mitigation, is missing on 
about 24\% of the embedded binaries. According to the breakdown results 
shown in Fig.~\ref{tab:mitigation-arch}, the lack of NX stack mostly 
happens to MIPS binaries. This is attributed to a MIPS-specific 
hardware restriction. The MIPS standards do not 
mandate the behaviors of Floating Point Unit (FPU) instructions. 
To normalize the behaviors of FPU operations, the Linux system
emulates certain FPU instructions and places the emulated code on 
the stack to execute~\cite{LinuxMIPS}. Thus, the stack is 
marked executable, and the situation only changes after a patch
in Linux kernel became available in 2016.

The above analysis shows a picture from the high level. 
To gain a deeper understanding, we break down the results 
from the dimensions of binary type, architecture, and vendor.

\subsubsection{Breakdown by Binary Type}
\label{subsec:binary-type}
Table~\ref{tab:elftype-mitigation} shows the results separately 
measured on different types of binary. The majority of embedded binaries
are dynamic libraries and dynamically linked executables 
(for simplicity, we call them dynamic executables), 
which adopt attack mitigations more often than their 
static counterparts. In particular, their adoption rates of 
Stack Canaries and RELRO are significantly higher. 
However, regardless of which type we consider, 
embedded binaries consistently have a lower adoption rate than 
desktop binaries on every mitigation. 

\subsubsection{Breakdown by Architecture}
Embedded devices use very diverse architectures, which 
affects the adoption of mitigations in different ways. For 
instance, as pointed out above, MIPS restricts the 
deployment of NX Stack. Inspired by this, we break down the measurement results based on
architectures. 


As shown in Fig.~\ref{tab:mitigation-arch}, embedded binaries running 
on every architecture have a relatively low rate of adopting attack 
mitigations (except NX Stack). However, the adoption rates do 
vary across architectures. ARM binaries constitute the largest group. 
They have a moderate level of adoption rate regardless of which 
mitigation we consider. MIPS binaries are the second largest
group. They, however, have the lowest adoption rate in 
nearly every mitigation. This is not surprising regarding 
NX Stack since MIPS imposes some 
hardware restrictions. The low adoption rates 
of other mitigations, in contrast, should reflect the 
choice of the vendors. 

Compared to MIPS binaries, x86 binaries have 
broader adoption of the mitigations. In particular, 
x86 binaries have the highest adoption 
rate of Stack Canaries among all the architectures. 
However, the observation may not reflect the general case.
Most of the x86 binaries
come from QNAP (see Table~\ref{tab:extracted-result}), 
which offers an 80.1\% adoption rate of Stack Canaries 
(see Table~\ref{tab:mitigation-result}). It is 
unclear whether the adoption rate will remain high
when more x86 binaries from other vendors are 
considered\footnote{We attempted to expand the dataset of x86 binaries,
but could not identify many other vendors using x86 architectures.}. 

AArch64 binaries and x64 binaries have relatively 
higher adoption rates in all the mitigations. In fact,
they present the highest adoption rates on RELRO, NX Stack, 
Fortify Source, and PIE. This is reasonable since 
the two architectures were released more recently. 
Binaries running on them tend to have newer building 
environments (\eg compiler and linker) and newer 
execution environments (\eg libraries and OS),
where the mitigations are better ready.

\subsubsection{Breakdown by Vendors}

Another interesting angle is to look at the differences across vendors, 
which helps answer questions such as which vendor offers the best 
mitigations. Related results have been presented in Table~\ref{tab:mitigation-result}.

The adoption rates of mitigations dramatically vary across vendors.
Vendors like AVM and Synology apply many attack mitigations
to most of their binaries. Others vendors like RouterTech and TomatoShibby
rarely adopt most mitigations. Zooming into individual mitigations, 
the difference is similarly intense. Consider Stack Canaries and Fortify Source 
as examples. AVM provides Stack Canaries to 81.5\% of its binaries, 
while 11 other vendors entirely omit Stack Canaries; Synology 
enables Fortify Source for 43.5\% of its binaries, but in contrast, 
27 other vendors only apply Fortify Source to less than 1\% of their
binaries. These differences reflect that the low adoptions rates 
of many vendors are a result of their (intentional or unintentional)
``choices'' instead of objective constraints.


Different vendors also have diverse ``preferences''. 
For instance, AVM prevalently applies Stack Canaries and RELRO while 
largely neglecting Fortify Source. On the contrary, 
Moxa and Synology prioritize Fortify Source but emphasize 
less on the AVM's preferred mitigations. Another generic 
observation is that more vendors have preferences for NX Stack 
and PIE. Presumably, this is because NX Stack and PIE have a lower cost, 
which is more amendable to embedded devices. 

\setlength\tabcolsep{4.6pt}
\begin{table}[t!]
  \centering
  \caption{Adoption of mitigations by different device types.}
  \vspace{-0.5em}
  \label{tab:device-category}
  \begin{tabular}{l c c c c c}
    \toprule
    \textbf{Device Type} & \textbf{Canary} & \textbf{RELRO} & \textbf{NX} & \textbf{Fortify} & \textbf{PIE}\\ \toprule
    Routers      & 4.1    & 8.0    & 58.5    & 0.1    &11.0\\  \hline
    WIFI Systems   & 0.4 & 1.4    & 35.7    & 0.2    & 11.4  \\  \hline
    Net-Switches   & 8.0 & 10.0    & 77.6    & 2.6    & 25.8  \\  \hline
    Modems   & 0 & 0    & 83.2    & 0    & 3.4  \\  \hline
    Net-Controllers   & 3.5 & 2.0    & 11.4    & 2.0    & 30.3  \\  \bottomrule
    Less-Networked    & 10.4 & 9.9        & 48.8         & 0.1         & 20.6         \\  \bottomrule
  \end{tabular}
  \vspace{-1em}
\end{table}
\setlength\tabcolsep{6pt}
\subsubsection{Breakdown by Device Types}
The adoption of attack mitigations may  
also be tied to the use scenarios of the devices. For instance, 
less-networked devices such as radio players have a lower risk 
of exploitation and thus, may skip the mitigations. 
To this end, we separately measured the attack mitigations
on different types of devices. As summarized in Table~\ref{tab:device-category}, 
the adoption rates of attack 
mitigations are not evidently disparate across device
types. All types of devices present an insufficient adoption of
Stack Canary (0-10.4\%), RELRO (0-10\%), and Fortify Source (0-2.4\%).
Network switches and modems demonstrate a relatively
higher adoption of NX Stack. This is not because of their types
but, instead, that network switches and modems are more ARM-based 
than MIPS-based. Moreover and very interestingly, the adoption rates
by internet-exposed devices are not higher than their less-networked 
counterparts. In summary, the results show no strong correlation
between mitigation adoption and device type.


\subsubsection{Mitigation Adoption and Vulnerability Presence} 
The presence of vulnerabilities is the key motivation 
of mitigations. This brings up two questions. 
First, \textit{are the low adoption rates of mitigations 
attributed to the lack the vulnerabilities}? Second, \textit{are the 
adoption rates higher on devices containing more vulnerabilities?} 
Accordingly, we perform a case study on embedded 
vulnerabilities. It shows that memory corruption vulnerabilities 
are common on embedded devices. For instance, two recently 
reported cases~\cite{realtek,AMNESIA33} both appear in
millions of embedded devices. This brings a negative answer to the
first question. In a follow-up step, we identified 1,360 binaries 
packaged as part of the Realtek SDK from 369 devices, which are affected
by one of the above vulnerabilities~\cite{realtek}. 
Compared to other binaries, these binaries present no broader 
adoption of the attack mitigations (Stack Canaries: 3.50\%; RELRO: 
1.60\%; NX Stack: 24.40\%; Fortified Source: 0.10\%; PIE: 2.50\%). 
This indicates a negative answer to the second question.

\vspace{0.5em}
\noindent
\fcolorbox{black}{babyblueeyes}{\parbox[c]{8.8cm}{
\textbf{Summary:} Embedded devices have a low rate of adopting 
user-space mitigations, despite these devices broadly have the 
needed supports. The low adoption rate is partially attributed 
to the restrictions of architectures and runtime environments, 
but it in general reflects the ``decisions'' of vendors.    
}}

\begin{figure}[!t]
        \centering
        \includegraphics[scale=0.19]{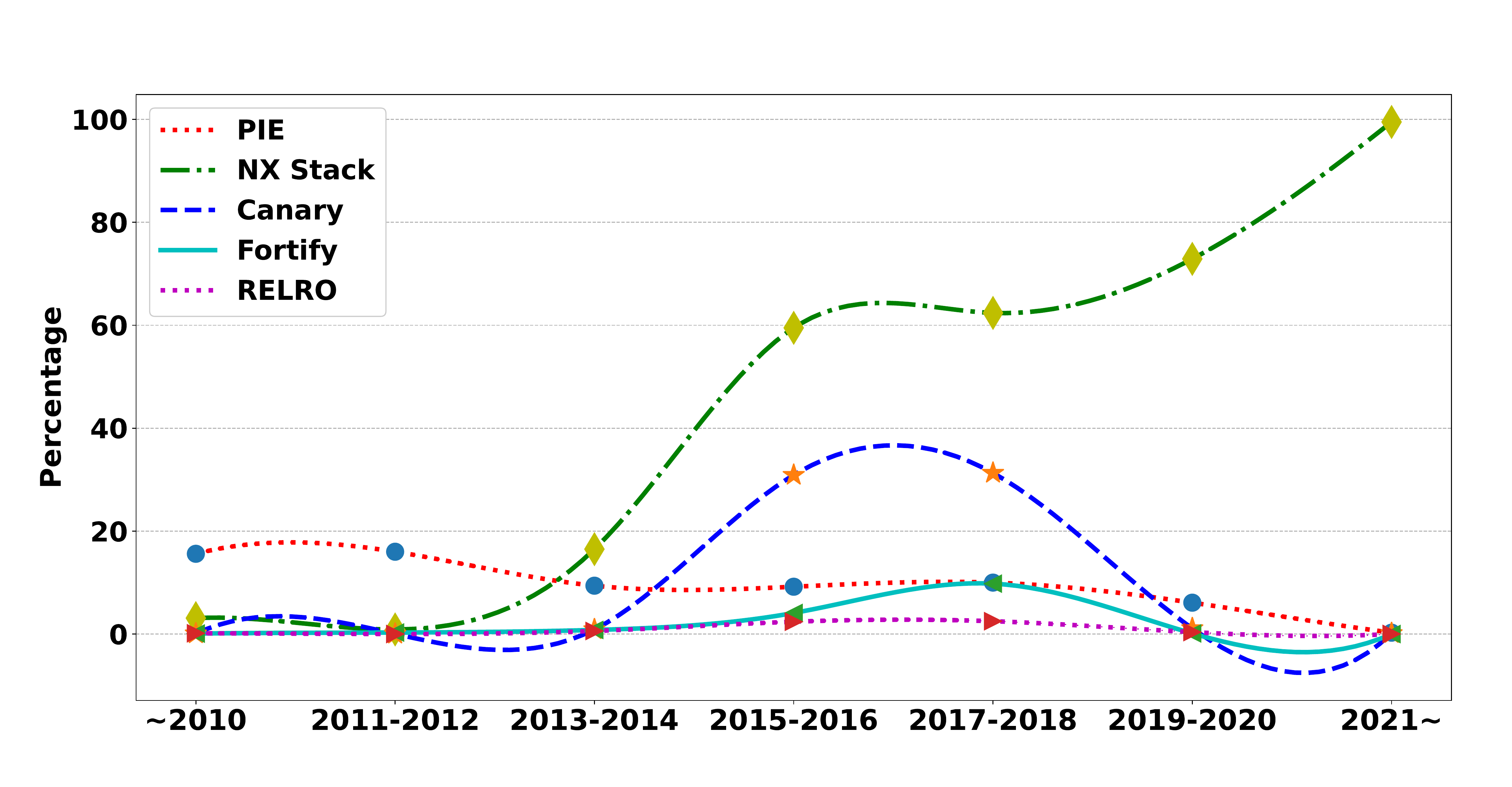}
        \vspace{-1em}
        \caption{Adoption rates of user-space mitigations across time. All binaries 
        released before 2010 are aggregated into \code{$\sim$2010}.}
        \label{figure:product-time}
        \vspace{-1.5em}
\end{figure}

\subsection{Changes over Time}
\label{subsec:user:change}

While the overall rates of adopting user-space mitigations
are not exciting, the situation might be improving. To verify this, 
we run a separate study to inspect the evolution over time. 


\begin{figure}[!t]
        \centering
        \includegraphics[scale=0.19]{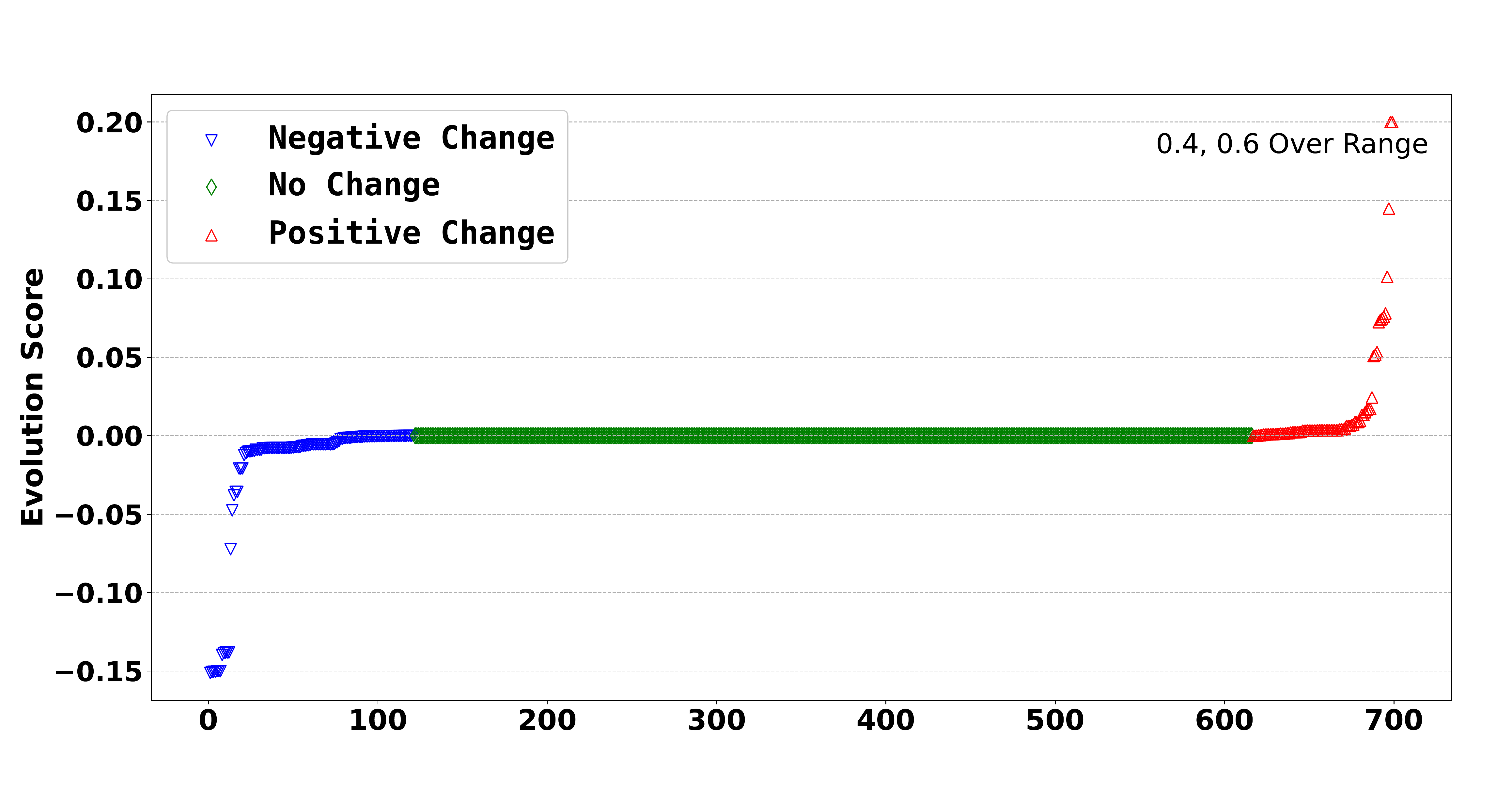}
        \vspace{-0.5em}
        \caption{Evolution score of individual firmware in the adoption of Stack Canaries. 
        Each point represents a firmware with multiple versions. The firmware 
        is sorted based on the evolution score. The two points marked \code{Over Range}
        at the upper left corner
        have evolution scores of \code{0.4} and \code{0.6}.}
        \label{figure:product-canary}
        \vspace{-1em}
\end{figure}

\subsubsection{The Overall Trend}
In this analysis, all binaries are grouped
based on the releasing time of their firmware\footnote{Firmware 
with unknown releasing time is excluded.}.
Specifically, binaries released in each period of two years
are grouped together. Each group is then
separately measured to understand their adoption of different mitigations. 

Fig.~\ref{figure:product-time} shows the changes over time. The adoption 
of NX Stack presents a consistent, positive trend. The adoption rate increased 
from nearly 0\% before 2010 to almost 100\% recently.
We believe a major reason is the increasing use of newer 
Linux kernels (see Fig.~\ref{figure:kernel-version-time}). 
The new kernels bring better supports for NX Stack, particularly
a patch to enable NX Stack in MIPS binaries~\cite{kernelgi68:online}.


The adoption rates of Stack Canaries and Fortify Source have a jump 
between 2015 and 2018. However, the jump may not represent
what happens in general. From 2016 to 2018, QNAP released a large number 
of binaries with broad adoption of Stack Canaries (see Table~\ref{tab:extracted-result} 
and Table~\ref{tab:mitigation-result}), pulling up the average adoption rate. 
In a similar way, Ubiquiti and Moxa raised the adoption rate 
of Fortify Source between 2016 and 2018. During 2018-2021 where 
these vendors released fewer binaries, the adoption rates
of both mitigations dropped. RELRO has a stable, low adoption rate 
in the past decade. In contrast,
PIE is more often adopted, but it presents a decreasing trend. 
Again, this decreasing trend may overfit the decisions of certain vendors.
For instance, OpenWrt released 190k binaries without PIE between 2020 
and 2021, resulting in the lowest adoption rate in the past decade. 

As described above, the overall trend can be significantly 
affected by the decisions of specific vendors at certain points
and thus, may not show the actual evolution. To this end, we perform 
two more fine-grained analyses.

\subsubsection{Evolution of Individual Firmware}
In our dataset, 699 firmware has multiple versions. 
The changes in mitigation adoption across different
versions are good indicators of evolution.
For each of the 699 firmware, we measure the \textit{evolution score},
namely the increase/decrease of adoption rate from its 
earliest version to its latest version. Fig.~\ref{figure:product-canary} 
shows the results of evolution score for Stack Canaries, and 
Fig.~\ref{fig:product-change-other} in the Appendix presents the 
results for other mitigations.

Most of the firmware presents no changes in adopting user-space 
mitigations. Among the few that indeed show changes, we observe 
both positive and negative trends. Consider Stack Canaries as an example. 
83 of the firmware has an increased adoption rate, while 121
has a reduced adoption rate. Breaking down the results to
individual vendors (see Table~\ref{tab:mitigation-vendor-range} 
in the Appendix), we observe that only Moxa offers a consistent, 
meaningful increase of adoption rate for all mitigations.
Most vendors either do not 
change or change positively for some mitigations but 
negatively for the others. Vendors like QNAP even 
consistently reduce the adoption rates of mitigations 
when upgrading their firmware.




\setlength\tabcolsep{6pt}
\begin{table}[!t]
  \centering
  \caption{Evolution of individual binaries in adopting attack mitigations. \code{No Change}, \code{Positive Change}, and \code{Negative Change} show the number of binaries without changes, with mitigation added, and with mitigation dropped.}
  \label{tab:overlap-binary-change}
  \vspace{-0.5em}
  \scriptsize
  \begin{tabular}{ c  c c c c c}
    \toprule
   \textbf{Category} & \textbf{Canary} & \textbf{RELRO} &\textbf{NX} & \textbf{Fortify} & \textbf{PIE} \\\toprule
    \code{No Change} & 278,877 & 278,992 & 278,375 & 279,434 & 278,674  \\ \hline
    \code{Positive Change} & 438 & 323 & 1,006 & 61 & 810  \\ \hline
    \code{Negative Change} & 284 & 284 & 218 & 104 & 115  \\ \bottomrule
  \end{tabular}
  \vspace{-1em}
\end{table}
\setlength\tabcolsep{6pt}

\subsubsection{Evolution of Individual Binaries}
The same binary may propagate across different versions of 
the firmware, which we call \textit{versioned binaries}. 
The change in mitigation adoption by versioned binaries is another
indicator of evolution. From our dataset, 
279k versioned binaries are identified\footnote{Binaries with 
the same name are considered as the same binary.} from 24 vendors. 
Each binary is then measured to see the change of mitigation adoption 
between its first and last versions. Table
~\ref{tab:overlap-binary-change} shows the aggregated 
results. Over 99.9\% of the versioned binaries present no 
changes in adopting the attack mitigations. Among the remaining, 
a significant portion shows negative changes. More details
for each vendor are presented in Table~\ref{tab:mitigation-overlap-binary} 
in the Appendix. Overall, no single vendor brings significantly 
broader adoption of mitigations to those binaries during upgrading.

\vspace{0.5em}
\noindent
\fcolorbox{black}{babyblueeyes}{\parbox[c]{8.8cm}{
\textbf{Summary:} There is no obvious evidence showing that
the adoption of user-space mitigations is improving.
}}

\begin{figure}[!t]
        \centering
        \includegraphics[scale=0.19]{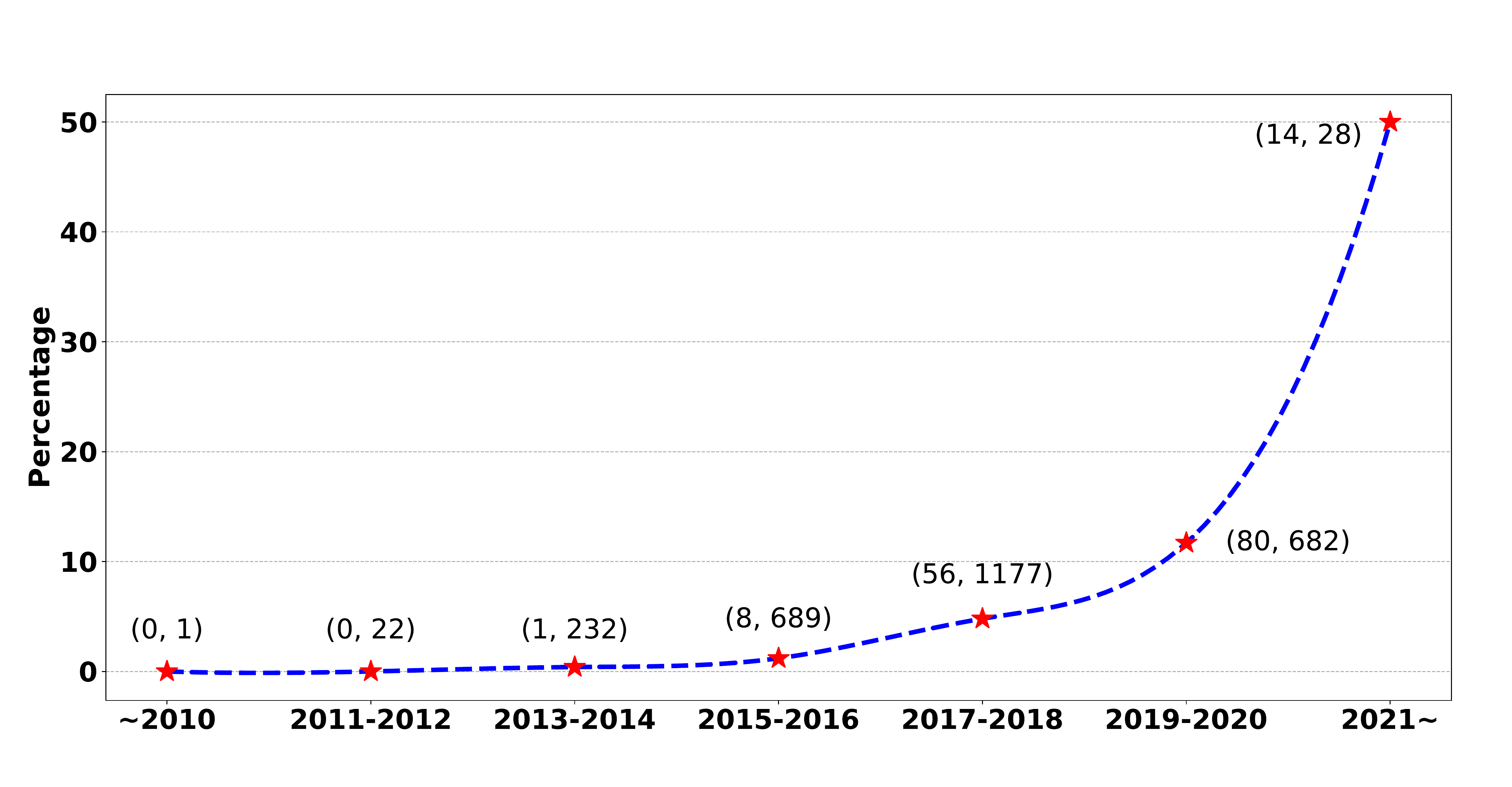}
        \vspace{-1.5em}
        \caption{Evolution of Stack Protector across time. The pair 
        of value \code{(x, y)} above each point means the total number of 
        applicable kernels is \code{x} and \code{y} of them are protected.
        The leap in 2021 may not reflect the general case since the total number of 
        kernels is too small.}
        \label{figure:kernel-canary-time}
        \vspace{-1.5em}
\end{figure}

\begin{table*}[t!]
  \centering
  \caption{Adoption of kernel-level mitigations. \code{Analyzed} indicates how many kernels can be analyzed to test the presence of each mitigation;  \code{Unsupported} means how many kernels have a version before the mitigation is integrated; \code{Protected} shows how many kernels adopt each mitigation.}
  \vspace{-0.5em}
  \label{tab:kernel-result} 
  \begin{tabular}{l c c c c c c c c}
    \toprule
    \textbf{Category} & \textbf{Total} & \textbf{Stack Protector} & \textbf{PXN} & \textbf{KASLR} & \textbf{FreeList} & \textbf{Usercopy}  & \textbf{Fortify} & \textbf{Kernel RWX}\\ \toprule
    \code{Analyzed} & 3,347 & 2,831  & 839  & 2,062 & 2,063 & 1,980  & 525  &    564  \\\hline
    \code{Unsupported} & - & 2,078  & 798  & 2,048 & 2,049 & 1,968  & 521  & 555 \\\hline
    \code{Protected} & - & 159  & 41  & 0 & 0 & 3  & 4  & 9     \\
    \bottomrule
  \end{tabular}
  \vspace{-1em}
\end{table*}
\section{Measuring Adoption of Kernel-Level Mitigations}
\label{sec:kernel-space}
 
From the collected firmware images, 7,977 Linux kernels 
are extracted. We run an analysis on the kernels to identify 
the mitigations in Table~\ref{tab:kernel-space-defense}. 
We only consider kernels that have \code{.config} files extracted
or can be converted to an ELF file as our identification
approach relies on that.
Given a target mitigation, we only include kernels built after 
the release of the mitigation and only include kernels using 
the desired architecture. Otherwise, the mitigation is 
certainly not adopted.




\begin{figure}[!t]
        \centering
        \includegraphics[scale=0.19]{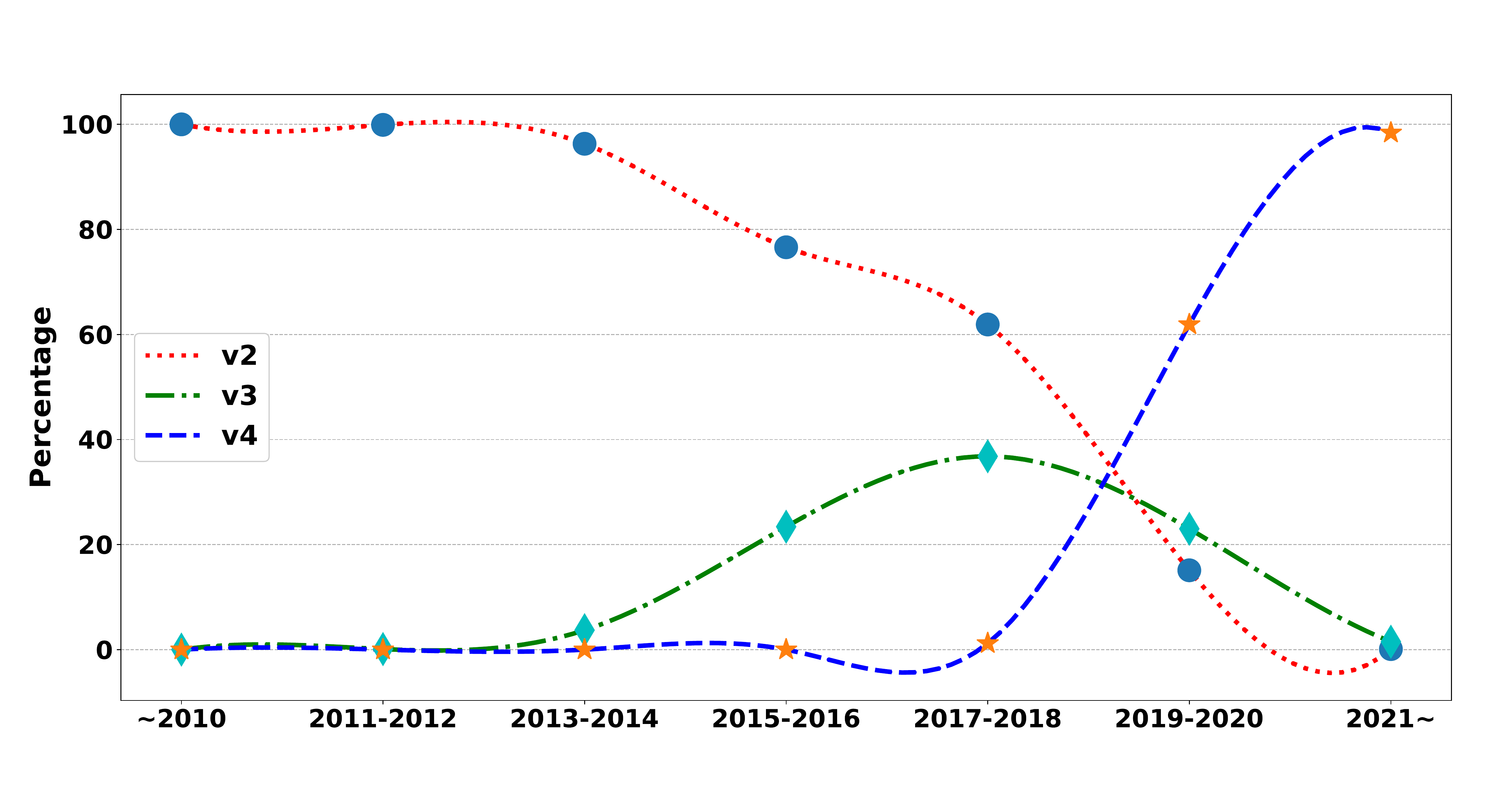}
        \vspace{-1.5em}
        \caption{Distribution of kernel versions across their building time. Subversions are disregarded and aggregated (\eg all v2.x.y are aggregated into v2). The leap of v4 between 2020 and 2021 is presumably attributable to a large number of Linux v4 firmware released by OpenWrt.}
        \label{figure:kernel-version-time}
        \vspace{-1.5em}
\end{figure}

\setlength\tabcolsep{3.9pt}
\begin{table}[t!]
  \centering
  \caption{Gap between the release time and the building time of kernels from our dataset (months). \textbf{Gap Range} shows the range of the gaps for all kernels from the same vendor.}
  \label{tab:kernel-timegap}
  \vspace{-0.5em}
  \begin{tabular}{l c c c}
    \toprule
    \textbf{Vendor} & \textbf{\# of Kernels} & \textbf{Gap Average} & \textbf{Gap Range} \\ \toprule
    CenturyLink & 1 & 81.0 & [81, 81] \\  \hline
    Phicomm & 3 & 56.7 & [32, 70]\\  \hline
    360 & 4 & 69.0 & [46, 78]\\  \hline
    Buffalo & 4 & 46.5 & [23, 86]\\  \hline
    Digi & 5 & 80.5 & [67, 102]\\  \hline
    Camius & 6 & 53.2 & [36, 64]\\  \hline
    Zyxel & 7 & 79.4 & [15, 101]\\  \hline
    Polycom & 16 & 115.5 & [69, 185]\\  \hline
    TENVIS & 27 & 79.4 & [71, 87]\\  \hline
    Mercury & 27 & 49.6 & [20, 64]\\  \hline
    Belkin & 57 & 63.1 & [15, 206]\\  \hline
    Linksys & 97 & 91.6 & [27, 187]\\  \hline
    Dlink & 114 & 71.1 & [18, 153]\\  \hline
    Netcore & 137 & 84.1 & [49, 168]\\  \hline
    RouterTech & 142 & 102.4 & [68, 132]\\  \hline
    Tenda & 142 & 81.5 & [28, 121]\\  \hline
    TRENDnet & 147 & 90.1 & [4, 177]\\  \hline
    Hikvision & 176 & 61.9 & [20, 109]\\  \hline
    Supermirco & 187 & 110.7 & [26, 174]\\  \hline
    TomatoShibby & 276 & 83.9 & [42, 93]\\  \hline
    Tp-Link-zh & 385 & 71.7 & [29, 138]\\  \hline
    ASUS & 415 & 78.9 & [29, 133]\\  \hline
    Ubiquiti & 426 & 77.2 & [16, 174]\\  \hline
    Tp-Link-en & 565 & 82.2 & [29, 139]\\  \hline
    Netgear & 1,263 & 68.0 & [12, 166]\\  \hline
    Openwrt & 2,755 & 40.2 & [36, 47]\\  \bottomrule
    
    \textbf{Average} & 284 & 65.1 & - \\  \bottomrule
  \end{tabular}
  \vspace{-1em}
\end{table}
\setlength\tabcolsep{6pt}

\subsection{Analysis of Results}
Table~\ref{tab:kernel-result} presents the results. 
Kernel-level mitigations are rarely adopted in embedded devices.
Stack Protector is applied the most, but sill only to 159 out 
of 2,831 kernels. The other mitigations have an adoption 
rate close to zero. In particular, KASLR and Freelist Randomization 
are not adopted at all. 

A major reason why kernel-level mitigations are missing is the vendors'
tendency to use old kernels. Fig.~\ref{figure:kernel-version-spread} 
in the Appendix shows the distribution of kernels across versions. 
Nearly half of the kernels have a version of v2.x, which was 
released almost 20 years ago. Looking closely at the kernels, 
we further find that they were mostly built years after the release,
as illustrated in Table~\ref{tab:kernel-timegap}. The average gap 
between the kernels' release time and building time is 
over 5 years (65.1 months). The gap for some kernels from 
Polycom and Linksys even goes over 15 years. 

A consequence of using old kernels is that the mitigations are unsupported.
As shown in Table~\ref{tab:kernel-result}, most kernels miss 
mitigations because they are not new enough to have the mitigations.
A very possible incentive for the vendors to use old kernels is reliability. 
When an older kernel runs well on the products,
it is often safer to continue using it since
an upgrade can easily introduce backward-incompatibility issues. 
To verify this intuition a bit, we again check the firmware 
with multiple versions but focus on the kernels this time.
As shown in Fig.~\ref{figure:kernel-version-change} in the Appendix, 
the vendors are in general ``reluctant'' to use newer kernels when upgrading 
their firmware, indirectly supporting our intuition.

\subsection{Changes over Time}
\label{subsec:kernel:change}

To understand the evolution of kernel-level mitigations, 
we perform an additional time series analysis. We only 
consider Stack Protector in this analysis because other 
mitigations have too few samples. The results,
presented in Fig.~\ref{figure:kernel-canary-time}, show a 
positive trend. The adoption rate of Stack Protector consistently
increases over the past decade. The driving force behind
the trend is mainly the upgrading of kernels. 
As demonstrated in Fig.~\ref{figure:kernel-version-time}, 
vendors are using more new kernels where Stack Protector
is more prevalently integrated. We envision this trend, 
in the longer term, 
will also benefit other mitigations.

\vspace{0.5em}
\noindent
\fcolorbox{black}{babyblueeyes}{\parbox[c]{8.8cm}{
\textbf{Summary}: Kernel-level mitigations are missing in 
embedded devices. A major reason is the vendors largely use
old kernels where the mitigations are not ready. 
Nonetheless, preliminary evidence shows positive changes are happening.
}}

\section{\enspace Discussion: Why Mitigations are Missing}
\label{sec:discussion}

The key takeaway of our study is that attack mitigations are
prevalently missing on embedded devices. Based on what we
have explained, the lack of kernel-level mitigations
is primarily caused by the excessive use of old kernels. 
In contrast, the problem of lacking user-space mitigations is
more complicated. Hardware/runtime restrictions (\eg the restriction
by MIPS on NX Stack) are undeniably a factor, but the 
primary reason should be the ``decision'' of vendors. The key 
question worth discussing here is \textit{why the vendors 
make such a decision}? 

Without comments from the vendors\footnote{We intentionally 
avoid interaction with vendors to prevent ethical issues.}, 
it is hard to get the exact answer to the above question. 
But throughout analysis of the commonalities shared by the firmware,
we are able to gain some observations that may help answer the question.

\subsection{Restrictions of Building Tools}

Vendors often rely on automated tools to build 
their embedded systems. \buildroot~\cite{buildroot}
is one of the most popular tools for Linux-based 
embedded systems. These automated tools may
delay the availability of attack mitigations for years. 
Consider \buildroot as an example. As shown in 
Table~\ref{tab:buildroot-mitigation} in the Appendix, 
it does not offer full support of Stack Canaries until 2013 
(8 years after the release of Stack Canaries). This similarly 
happens to other mitigations. In this regard,
the use of automated tools like \buildroot (in particular 
the older versions) defers or even prevents the adoption 
of attack mitigations.

\begin{figure}[!t]
        \centering
        \includegraphics[scale=0.195]{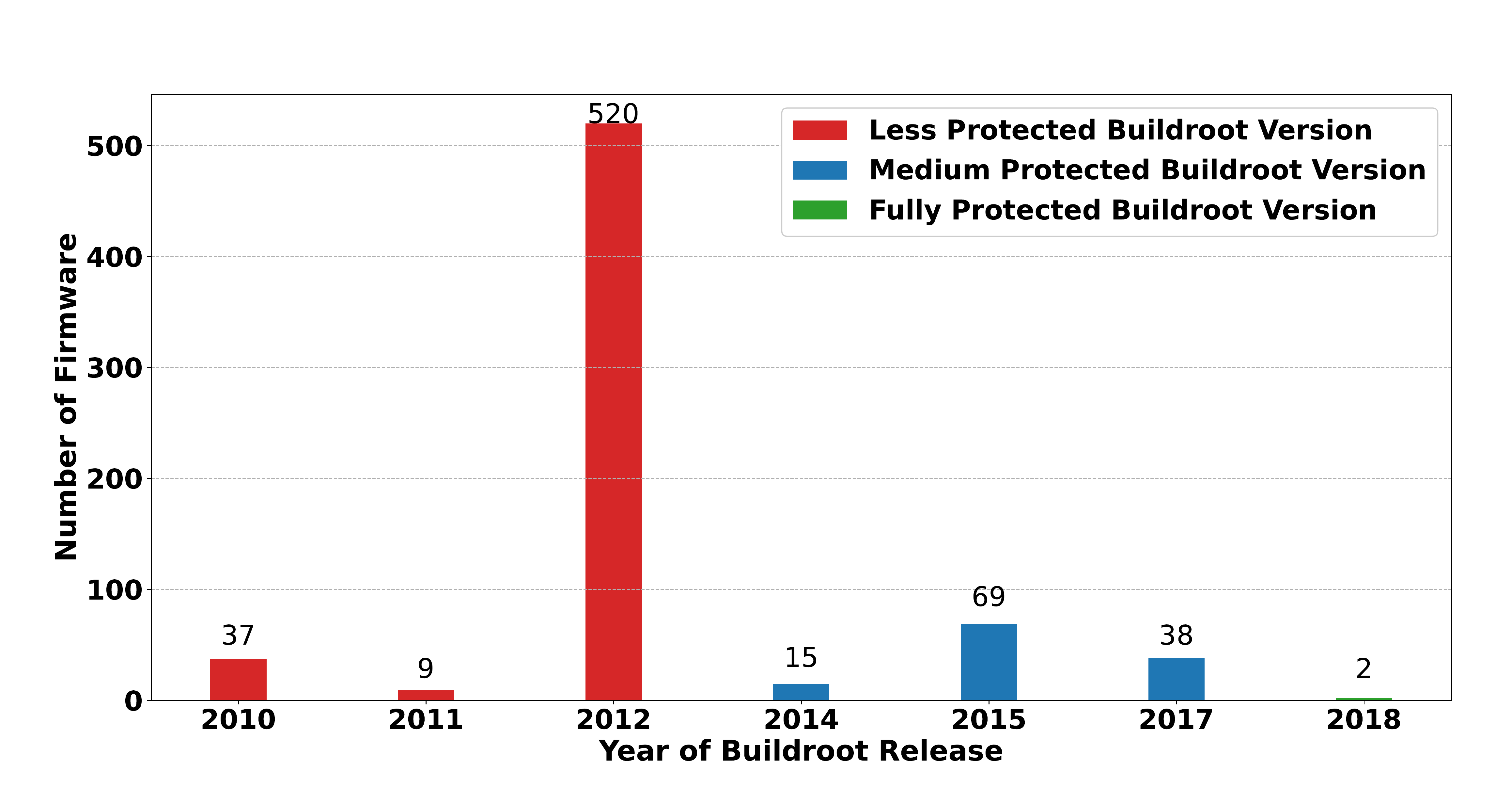}
        \vspace{-1em}
      \caption{Distribution of firmware based on the version of \buildroot they are built with.}
      \label{fig:buildroot-version}
      \vspace{-1.5em}
\end{figure}

\setlength\tabcolsep{3.8pt}
\begin{table*}[h]
  \centering
  \caption{Number of unique binaries from each vendor (k). The uniqueness of a binary is defined by its MD5.}
  \vspace{-0.5em}
  \label{tab:overlap-binary}
  \footnotesize
  \begin{tabular}{ l| c| c| c| c| c| c| c| c| c| c| c| c| c| c| c| c| c| c| c| c| c| c| c| c| c| c| c| c| c| c| c| c| c| c| c}
 \toprule

\rotatebox[origin=c]{90}{\textbf{Vendor}} & \rotatebox[origin=c]{90}{TENVIS}    &  \rotatebox[origin=c]{90}{Haxorware} & \rotatebox[origin=c]{90}{Cerowrt} & \rotatebox[origin=c]{90}{AT\&T} & \rotatebox[origin=c]{90}{Camius} & \rotatebox[origin=c]{90}{GOCloud} & \rotatebox[origin=c]{90}{Actiontec} & \rotatebox[origin=c]{90}{Buffalo} & \rotatebox[origin=c]{90}{360} & \rotatebox[origin=c]{90}{Phicomm} & \rotatebox[origin=c]{90}{Mercury} & \rotatebox[origin=c]{90}{Zyxel} & \rotatebox[origin=c]{90}{CenturyLink}  & \rotatebox[origin=c]{90}{Digi} & \rotatebox[origin=c]{90}{Supermirco} & \rotatebox[origin=c]{90}{AVM} & \rotatebox[origin=c]{90}{MikroTik} & \rotatebox[origin=c]{90}{OpenWrt} & \rotatebox[origin=c]{90}{Belkin} & \rotatebox[origin=c]{90}{NETCore} & \rotatebox[origin=c]{90}{RouterTech} &\rotatebox[origin=c]{90}{Dlink} &\rotatebox[origin=c]{90}{Tenda} &\rotatebox[origin=c]{90}{TomatoShibby} &\rotatebox[origin=c]{90}{Linksys}&\rotatebox[origin=c]{90}{TRENDnet}&\rotatebox[origin=c]{90}{Moxa}&\rotatebox[origin=c]{90}{QNAP}&\rotatebox[origin=c]{90}{Tp-Link-zh }&\rotatebox[origin=c]{90}{Ubiquiti}&\rotatebox[origin=c]{90}{Tp-Link-en} &\rotatebox[origin=c]{90}{NETGEAR}&\rotatebox[origin=c]{90}{ASUS}&\rotatebox[origin=c]{90}{Synology}&\rotatebox[origin=c]{90}{\textbf{Average}}\\ \toprule

\rotatebox[origin=c]{90}{\textbf{ Total }} & \rotatebox[origin=c]{90}{0.9} & \rotatebox[origin=c]{90}{0.2} & \rotatebox[origin=c]{90}{0.2} & \rotatebox[origin=c]{90}{0.4} & \rotatebox[origin=c]{90}{0.6} & \rotatebox[origin=c]{90}{0.5} & \rotatebox[origin=c]{90}{0.9} & \rotatebox[origin=c]{90}{0.4} & \rotatebox[origin=c]{90}{0.5} & \rotatebox[origin=c]{90}{0.5} & \rotatebox[origin=c]{90}{1.9} & \rotatebox[origin=c]{90}{1.5} & \rotatebox[origin=c]{90}{0.8} & \rotatebox[origin=c]{90}{1.5} & \rotatebox[origin=c]{90}{1.5} & \rotatebox[origin=c]{90}{5.0} & \rotatebox[origin=c]{90}{4.3} & \rotatebox[origin=c]{90}{191.2}& \rotatebox[origin=c]{90}{7.9} & \rotatebox[origin=c]{90}{10.2} & \rotatebox[origin=c]{90}{25.8} & \rotatebox[origin=c]{90}{15.9} & \rotatebox[origin=c]{90}{33.6} & \rotatebox[origin=c]{90}{127.8} & \rotatebox[origin=c]{90}{17.1} & \rotatebox[origin=c]{90}{15.3} & \rotatebox[origin=c]{90}{32.0} & \rotatebox[origin=c]{90}{296.0} & \rotatebox[origin=c]{90}{65.7} & \rotatebox[origin=c]{90}{204.7} & \rotatebox[origin=c]{90}{76.3} & \rotatebox[origin=c]{90}{173.9} & \rotatebox[origin=c]{90}{273.2} & \rotatebox[origin=c]{90}{1375.4}& \rotatebox[origin=c]{90}{87.2}\\\hline

\rotatebox[origin=c]{90}{\textbf{ Unique }}& \rotatebox[origin=c]{90}{0.1} &  \rotatebox[origin=c]{90}{0.2} & \rotatebox[origin=c]{90}{0.2} &  \rotatebox[origin=c]{90}{0.2} &  \rotatebox[origin=c]{90}{0.2} & \rotatebox[origin=c]{90}{0.3} &   \rotatebox[origin=c]{90}{0.3} &  \rotatebox[origin=c]{90}{0.3} &  \rotatebox[origin=c]{90}{0.4} &  \rotatebox[origin=c]{90}{0.5} & \rotatebox[origin=c]{90}{0.5} & \rotatebox[origin=c]{90}{0.5} &  \rotatebox[origin=c]{90}{0.7}  & \rotatebox[origin=c]{90}{0.9} & \rotatebox[origin=c]{90}{0.9}  &    \rotatebox[origin=c]{90}{1.8} &  \rotatebox[origin=c]{90}{2.1} &  \rotatebox[origin=c]{90}{2.9} & \rotatebox[origin=c]{90}{3.4} & \rotatebox[origin=c]{90}{3.7} & \rotatebox[origin=c]{90}{4.0} & \rotatebox[origin=c]{90}{4.7} &  \rotatebox[origin=c]{90}{5.8} & \rotatebox[origin=c]{90}{5.8}&\rotatebox[origin=c]{90}{7.0} & \rotatebox[origin=c]{90}{7.8} & \rotatebox[origin=c]{90}{12.8} &  \rotatebox[origin=c]{90}{12.8} & \rotatebox[origin=c]{90}{15.8}&  \rotatebox[origin=c]{90}{20.5} & \rotatebox[origin=c]{90}{23.6} &  \rotatebox[origin=c]{90}{29.2}&  \rotatebox[origin=c]{90}{29.7} & \rotatebox[origin=c]{90}{64.9}& \rotatebox[origin=c]{90}{7.8}\\\hline

\rotatebox[origin=c]{90}{\textbf{ Ratio (\%) }} & \rotatebox[origin=c]{90}{11.1} & \rotatebox[origin=c]{90}{100} & \rotatebox[origin=c]{90}{100} & \rotatebox[origin=c]{90}{50.0} & \rotatebox[origin=c]{90}{33.3} & \rotatebox[origin=c]{90}{60.0} & \rotatebox[origin=c]{90}{33.3} & \rotatebox[origin=c]{90}{75.0} & \rotatebox[origin=c]{90}{80.0} & \rotatebox[origin=c]{90}{100} & \rotatebox[origin=c]{90}{26.3} & \rotatebox[origin=c]{90}{33.3} & \rotatebox[origin=c]{90}{87.5} & \rotatebox[origin=c]{90}{60.0} & \rotatebox[origin=c]{90}{60.0} & \rotatebox[origin=c]{90}{36.0} & \rotatebox[origin=c]{90}{48.8} & \rotatebox[origin=c]{90}{1.5}& \rotatebox[origin=c]{90}{43.1} & \rotatebox[origin=c]{90}{36.3} & \rotatebox[origin=c]{90}{15.5} & \rotatebox[origin=c]{90}{29.6} & \rotatebox[origin=c]{90}{17.3} & \rotatebox[origin=c]{90}{4.5} & \rotatebox[origin=c]{90}{40.9} & \rotatebox[origin=c]{90}{50.9} & \rotatebox[origin=c]{90}{40.0} & \rotatebox[origin=c]{90}{4.3} & \rotatebox[origin=c]{90}{24.1} & \rotatebox[origin=c]{90}{10.1} & \rotatebox[origin=c]{90}{30.9} & \rotatebox[origin=c]{90}{16.8} & \rotatebox[origin=c]{90}{10.9} & \rotatebox[origin=c]{90}{4.7}& \rotatebox[origin=c]{90}{8.9}
    \\ \bottomrule
  \end{tabular}
  \vspace{-1.5em}
\end{table*}

\setlength\tabcolsep{6pt}

To gain a more concrete understanding, we zoom into the firmware
in our dataset generated with \buildroot. In total, there 
are 690 of them. As shown in Fig.~\ref{fig:buildroot-version}, 
most firmware was built with \buildroot released at 
or before 2012, when no mitigation was supported.
In result, all binaries built together 
with the firmware would carry zero attack mitigations.

But should we only blame the automated tools? The answer 
is clearly no. The vendors largely use 
old versions of \buildroot even newer versions
are available for years, just like what they did to 
Linux kernels. Table~\ref{tab:buildroot-timegap} shows
the gap between the release time of \buildroot and 
the use time by vendors. On average, the gap is 5 years.
Even the smallest gap is about 2.5 years. This amplified the delay of 
the availability of attack mitigations. 
So essentially, the lateness in integrating attack
mitigations by the automated tools and the use of 
older automated tools by embedded vendors together lead
to an barrier to the adoption of attack mitigations.

\subsection{Massive Reuse of Binaries}
We find that embedded vendors largely reuse binaries across products.
First, the same vendor often runs the same group of binaries 
on different devices. Table~\ref{tab:overlap-binary}
shows the number of unique binaries from 
each vendor. On average, only 8.9\% of the binaries 
are unique. In other words, the same binary is reused for 
11 firmware. Second, the binaries can also frequently propagate
across vendors. Fig.~\ref{fig:overlap-vendors} in the Appendix 
presents the number of unique binaries shared by multiple
vendors. Over 8,000 binaries are used by 2 vendors 
and some binaries are even used by 10 vendors. 
The heatmap in Fig.~\ref{tab:overlap-heatmap} in the Appendix 
gives more details about how frequently vendors
borrow binaries from each other. 

Reusing binaries across products or even vendors is 
understandable since these binaries have proven reliability.
But how exactly the reuse of binaries affects attack mitigations? 
To answer this question, we measure the adoption 
of attack mitigations in binaries reused by multiple vendors. 
In total, there are 11,232 such binaries. The binaries 
present a significantly lower adoption rate in most of the 
mitigations, compared to the results with all binaries 
considered (compare Table~\ref{tab:overlap-mitigation-result} 
and Table~\ref{tab:mitigation-result}). That means 
the propagation of those binaries brings harm to the 
overall adoption of attack mitigations. Even more worrisome 
is that the harm will continue until those binaries get rebuilt 
and redistributed. 

\begin{table}[t!]
  \centering
  \caption{Cost of attack mitigations on SPEC CPU2006. From left to right, the columns show the \textbf{accumulative overhead} after we enable the mitigations, one after another.}
  \label{tab:spec-overhead}
  \vspace{-0.5em}
  \begin{tabular}{c c c c c c}
    \toprule
    \textbf{Overhead} & \textbf{NX} & \textbf{Canary} & \textbf{PIE} & \textbf{RELRO} & \textbf{Fortify}\\ \toprule
    Storage   & 0 & 6.7\%    & 11.5\%    & 17.3\%    & 17.3\%  \\  \hline
    Memory    & 0 & 0        & 0         & 0         & 0         \\\hline
    Runtime & 0 & 6.6\%    & 8.45\%    & 10.7\%    & 10.9\%    \\  \bottomrule
  \end{tabular}
  \vspace{-1em}
\end{table}
\setlength\tabcolsep{6pt}
\subsection{Cost of the Mitigations}
The mitigations can bring extra cost, becoming a 
possible reason affecting their adoptions. To quantify the cost, we 
perform an evaluation of the storage/memory/performance overhead
of user-space mitigations on SPEC CPU2006, using a Raspberry PI-4B board 
as the device (Broadcom BCM2711 Quad-core Cortex-A72 with 8GB 
LPDDR4-3200 SDRAM). 

Table~\ref{tab:spec-overhead} shows the evaluation results. Overall, applying the mitigations together has no observable
overhead on memory usage but introduces a 10.9\% and 17.3\% overhead on  performance and binary size. Specifically, mitigations including Stack Canaries, PIE, and RELRO incur observable overhead in both performance
and binary size. Fortify Source, in contrast, brings a lightweight performance overhead without increasing the binary size. We believe
these types of overhead may impede the vendors from adopting
the mitigations. However, whether that indeed happens needs confirmation with the vendors, which we intentionally avoid for ethical concerns. 

\vspace{0.5em}
\noindent
\fcolorbox{black}{babyblueeyes}{\parbox[c]{8.8cm}{
\textbf{Summary}: Utilization of old building tools and massive reuse of existing binaries are contributors 
to the lack of attack mitigations in embedded binaries. Cost of the attack mitigations may also potentially 
impede their adoption.}}


\section{\enspace \enspace Threats to Validity}
\label{sec:threat}


\subsection{Representativeness of Dataset}
Similar to other sampling-based studies, our study can
present findings biased towards the collected dataset. 
We considered this threat and extended two efforts to reduce the threat. 
First, we attempted to cover 
all vendors that are popular or included in previous 
studies. Second, we enumerate the firmware images that are publicly
available from each vendor. In the end, we collected 
18k firmware images from 38 vendors. The number of 
firmware images and the list of vendors are comparable
to existing large-scale studies on embedded security~\cite{feng2016scalable,david2018firmup,
costin2014large,chen2021sharing,chen2016towards}.


\begin{table}[t!]
  \centering
  \caption{Gap between the release time and the use time of \buildroot (months). \textbf{Gap Range} shows the range of the gaps for all firmware from the same vendor.}
  \label{tab:buildroot-timegap}
  \vspace{-0.5em}
  \begin{tabular}{l c c c}
    \toprule
    \textbf{Vendor} & \textbf{Number} & \textbf{Gap Average} & \textbf{Gap Range} \\ \toprule
    360 & 1 & 50.0 & [50, 50] \\  \hline
    Belkin & 2 & 31.5 & [29, 32] \\  \hline
    Linksys & 20 & 73.0 & [40, 104] \\  \hline
    Netcore & 8 & 72.4 & [44, 98] \\  \hline
    TRENDnet & 9 & 48.9 & [8, 98] \\  \hline
    Tenda & 57 & 47.6 & [17, 104] \\  \hline
    TomatoShibby & 23 & 32.0 & [26, 54] \\  \hline
    Tp-Link-zh & 71 & 52.6 & [29, 98] \\  \hline
    Tp-Link-en & 158 & 63.7 & [28, 89] \\  \hline
    Dlink & 6 & 58.9 & [40, 80] \\  \hline
    NETGEAR & 259 & 67.9 & [22, 104] \\  \hline
    ASUS & 59 & 51.0 & [31, 66] \\  \hline
    Hikvision & 15 & 33.1 & [24, 38] \\  \hline
    Ubiquiti & 2 & 64.0 & [62, 66] \\  \hline
    \textbf{Average} & 49.3 & 60.0 & - \\  \bottomrule
  \end{tabular}
  \vspace{-1em}
\end{table}

\subsection{Imbalance in Dataset}
The dataset we collected is not perfectly balanced, which may 
harm our findings. First, not every vendor has the 
same amount of data involved. Vendors like OpenWrt and NETGEAR
contributed a large portion of the data, while other vendors like 
Cerowrt and Haxorware only provided a tiny part. 
Therefore, our generic findings
may overfit the dominant vendors. To this end, 
we break down the results to each vendor in most analyses, 
which helps raise awareness of the overfitting results. Overall, 
our generic findings broadly align with the breakdown 
results. Second, given a mitigation, the applicable
data samples are not evenly distributed over time. 
This can create abnormal points in the analysis of 
evolution trends. For instance, QNAP released a large number
of binaries with Stack Canaries from 2016 to 2018, 
causing a sudden leap in the evolution trend 
(recall \S\ref{subsec:user:change}). To mitigate the threat, 
we revisit all the points that appear to be outliers, 
followed by clarifying the impact of
data imbalance (see \S\ref{subsec:user:change} and 
\S\ref{subsec:kernel:change}). 


\subsection{Reliability of Mitigation Identification}

First, obfuscations can affect our identification of attack mitigations.
For instance, encoding strings can mislead our detection of Stack Canaries and 
Fortify Source. In contrast, destroying symbols can disrupt 
the detection relying on indicator functions. However, we envision 
that obfuscations should not have affected our study. 
First, we manually checked many binaries and did not observe obfuscations. 
Second, the firmware we collected is exclusively from mainstream, 
benign vendors who have fewer motivations to obfuscate the code.


Second, we rely on static approaches to 
identify mitigations, which can raise two problems. First, 
some binaries may not be used at all, so their results 
do not matter. We did not exclude such binaries. In 
this regard, our study covers a superset of truly 
security-relevant binaries. Second, some mitigations can be 
affected by runtime configurations. For instance, KASLR can be 
disabled by setting the \code{nokaslr} parameter when booting 
the kernel~\cite{nokaslr}. We, without knowing what happens
at runtime, cannot exclude such mitigations.
We believe the two problems should not affect our findings much. 
Based on our study, the adoption rates of attack mitigations
are extremely low, which shall still hold even considering a subset 
of the data and excluding the falsely identified mitigations.

Third, the tools we reuse to help identify may 
have reliability issues. For instance, \firmadyne can extract 
incomplete kernel data, and \vmlinuxtool can miss recovering 
kernel functions. Both will hurt our identification of 
kernel-level mitigations. We realize this threat, but we 
consider the reliability of existing tools out of this paper's
scope.





\section{Related Work}
\label{sec:related}

\subsection{Study of Threats to Embedded Devices} 

Past research has launched many attempts trying to 
understand the threats faced by embedded devices. 
Alrawi~\etal~\cite{alrawi2019sok} propose a modeling methodology
to systematize the security of home-based IoT devices from 
the dimensions of attack vectors, mitigations, and stakeholders. 
They further evaluated 45 open-source or on-market home-based 
IoT devices, which confirms security issues discussed by their study.
Inspired by the study and the evaluation, the authors eventually 
propose a list of mitigations to address the related security issues.
Falling into the category of studying security mitigations, 
they focus more on discussing the possible mitigations
against the attack vectors instead of the adoption of 
the mitigations in the wild. 

Costin~\etal~\cite{costin2014large} present a large-scale study 
to measure security vulnerabilities in 32k firmware images running 
on embedded devices. They designed and implemented a distributed
architecture to statically measure similarities between firmware
images, using a correlation engine. 
The study discovered 38 unknown vulnerabilities from 693 firmware 
images. The study also unveils that vulnerabilities from known affected
devices can ``propagate'' to other devices. Similarly, Feng~\etal~\cite{feng2016scalable} propose and implement Genius, a bug search engine based on features in the control-flow graph.  
Evaluating Genius on a dataset of 33,045 firmware, 
the authors discovered 38 potentially vulnerable firmware images 
from 5 vendors. The two studies and ours all rely on static 
analysis on a large corpus of firmware to understand embedded
security. However, they focus on vulnerabilities while 
we focus on attack mitigations. 

Chen~\etal~\cite{chen2016towards} develop \firmadyne, 
an automated, dynamic firmware analysis system, to 
run firmware binaries through full system emulation 
and an instrumented kernel. Leveraging \firmadyne, 
they run 74 exploits on 9,486 firmware images. 
The results unveil that firmware images are largely 
vulnerable to existing vulnerabilities and exploits:
887 of the firmware images supporting at least 89 distinct
products can be affected by one or more of the exploits.
Our study complements this work by understanding 
the adoption of mitigations against those exploits.



\subsection{Study of Attack Mitigations for Embedded Devices} 
There also exist studies on attack mitigations in embedded devices. 
The most closely related one to our study is presented in~\cite{buildsafety}. 
The authors evaluate the availability of ASLR, Non-executable Stack, RELRO, 
and Stack Canaries on 28 popular home routers with either ARM or MIPS architecture. 
The study presents some similar observations to ours. For instance, 
it finds that the adoption rate of Stack Canaries is extremely low, 
and NX Stack is more likely to be applied to ARM binaries than MIPS
binaries. Compared to this study, ours has a much larger scope
and a much higher depth, bringing more systematic insights towards 
improving the situation. 

Other studies in this line focus more on 
the challenges in applying attack mitigations to embedded devices. 
Thompson and Zatko~\cite{LinuxMIPS} explore
why MIPS devices usually miss applying NX Stack, as we 
explained in \S\ref{sec:user-space}. Abbasi~\etal~\cite{abbasi2019challenges} investigate 
the challenges faced by embedded devices to adopt attack mitigations.
They found that many embedded devices, particularly the low-end
ones, often lack the hardware and OS support 
needed by the mitigations. In our study, we concentrate on Linux-based
devices where the hardware and the OS are less restricted. We
aim to understand that, when objective restrictions like those 
discussed in~\cite{LinuxMIPS} do not exist, how often the mitigations
will be adopted.

\subsection{Tools for Mitigation Measurement} 
At the time we conduct the study, many existing tools, 
including \checksec~\cite{checksec}, \hardeningcheck~\cite{hardening-check}, 
 and \pwntools~\cite{pwntools}, 
can help detect user-space mitigations.
\checksec~\cite{checksec} is a bash script designed to 
test standard security properties of ELF files.
It additionally provides the feature of 
detecting kernel-level mitigations in running systems.
\hardeningcheck~\cite{hardening-check} is another tool
providing similar features as \checksec. \pwntools~\cite{pwntools} is a CTF framework 
and exploit development library. It provides functionality to 
check the status of the above security features applied in 
ELF binaries. We follow these tools to develop many of our 
detection strategies, but also extend them. 
First, we extend the mitigation detection to handle cases like 
Stack Canaries in statically linked, stripped binaries. 
Second, we add new supports to detect mitigations applied 
in the kernel without running it in the actual device.

\section{Conclusion}
\label{sec:conclusion}

This paper measures the adoption of standard attack mitigations 
in embedded devices. It shows that attack mitigations are largely 
missing even on devices where the needed hardware/OS supports 
are fully available. The findings also complement previous research
that ties the absence of mitigations to the lack of hardware/OS 
supports. By inspecting the evolution over time, the study unveils
that the situation does not improve in the past decade, casting 
a worrisome prediction about the upcoming future where embedded devices
will explode. On the positive side, the study identifies a set of 
doings hurting the adoption of attack mitigations, which bring 
insights towards improving the current practice.

\section*{Acknowledgments}
\label{sec:acks}

We thank our shepherd Avesta Sasan and the anonymous reviewers for their feedback. This project was supported by National Science Foundation (Grant\#: CNS-2031377), Office of Naval Research (Grant\#: N00014-17-1-2787; N00014-17-1-2788), and DARPA (Grant\#: D21AP10116-00). Any opinions, findings, and conclusions or recommendations expressed in this paper are those of the authors and do not necessarily reflect the views of the funding agency.


\bibliographystyle{IEEEtranS}

{\footnotesize
\bibliography{ref}}

\appendix

\begin{figure*}[ht]
        \centering
        \includegraphics[scale=0.55]{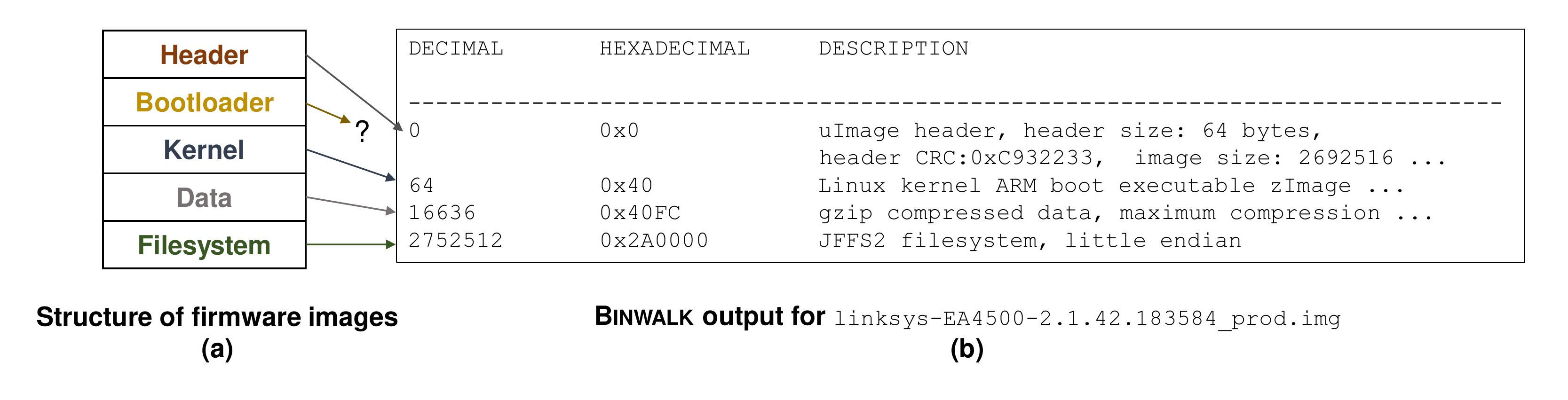}
        \caption{An example of Linux-based firmware image.}
        \label{tab:fw_img}
\end{figure*}

\begin{figure*}[t!]
  \centering
\begin{lstlisting}[style={arm}, escapechar=`, caption={The \code{stack\_chk\_fail} function in a statically-linked binary.}, label=lst:static_canary]
;function start
30: ldr     r3                  , data_101a0
34: ldr     r4                  , data_101a4
38: push    {r0, r1, r2, lr}
3c: ldr     r6                  , data_101a8
40: add     r4                  , pc        ,  r4   ; {data_6a28f, `\textcolor{red}{*** stack smashing detected ***}`:...}
44: ldr     r5                  , [pc, r3]
48: mov     r0                  , r4                ; {data_6a28f, `\textcolor{red}{*** stack smashing detected ***}`:...}
4c: bl      sub_10110
......
9c: bl      sub_64d94
;function end
\end{lstlisting}

\end{figure*}

\begin{figure*}[ht]
	\centering
	\includegraphics[width=\linewidth]{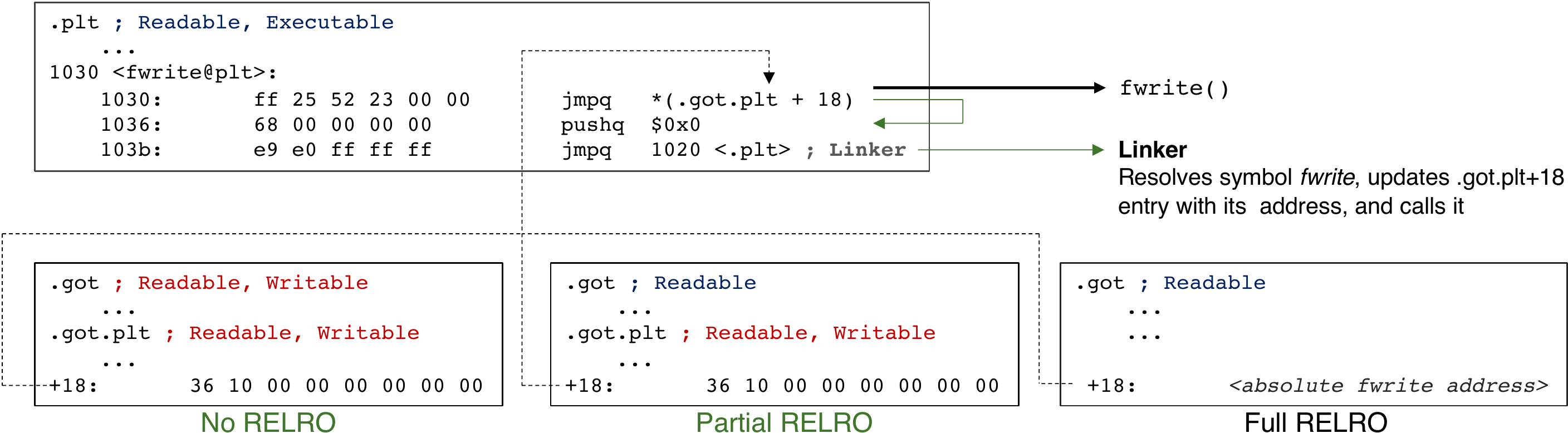}
	\caption{Binary with no/partial/full RELRO.}
	\label{fig:relro}
\end{figure*}

\begin{figure*}[ht]
	\centering
	\includegraphics[width=\linewidth]{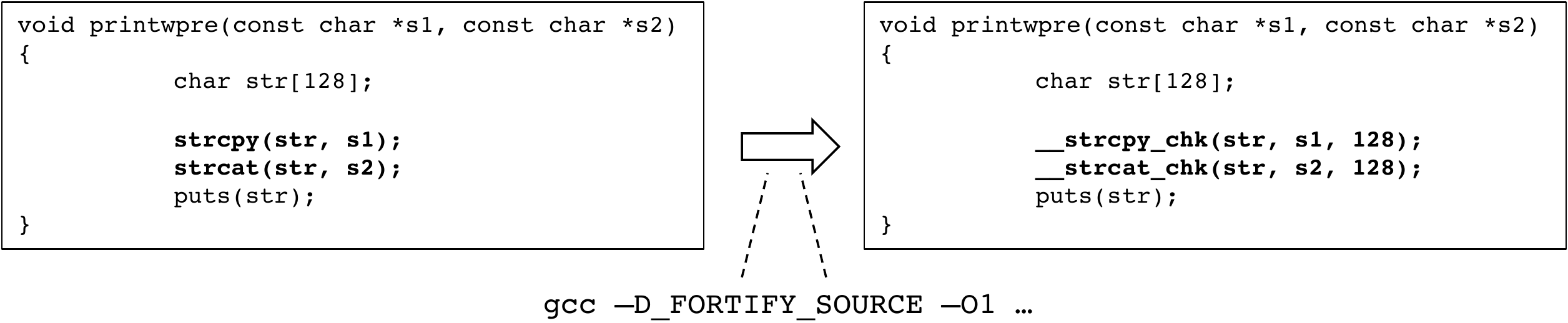}
	\caption{Replacement of dangerous libc functions with safer versions by Fortify Source.}
	\label{fig:fortify}
\end{figure*}

\begin{figure*}[ht]
        \centering
    \begin{subfigure}[b]{0.45\textwidth}
        \centering
        \includegraphics[scale=0.18]{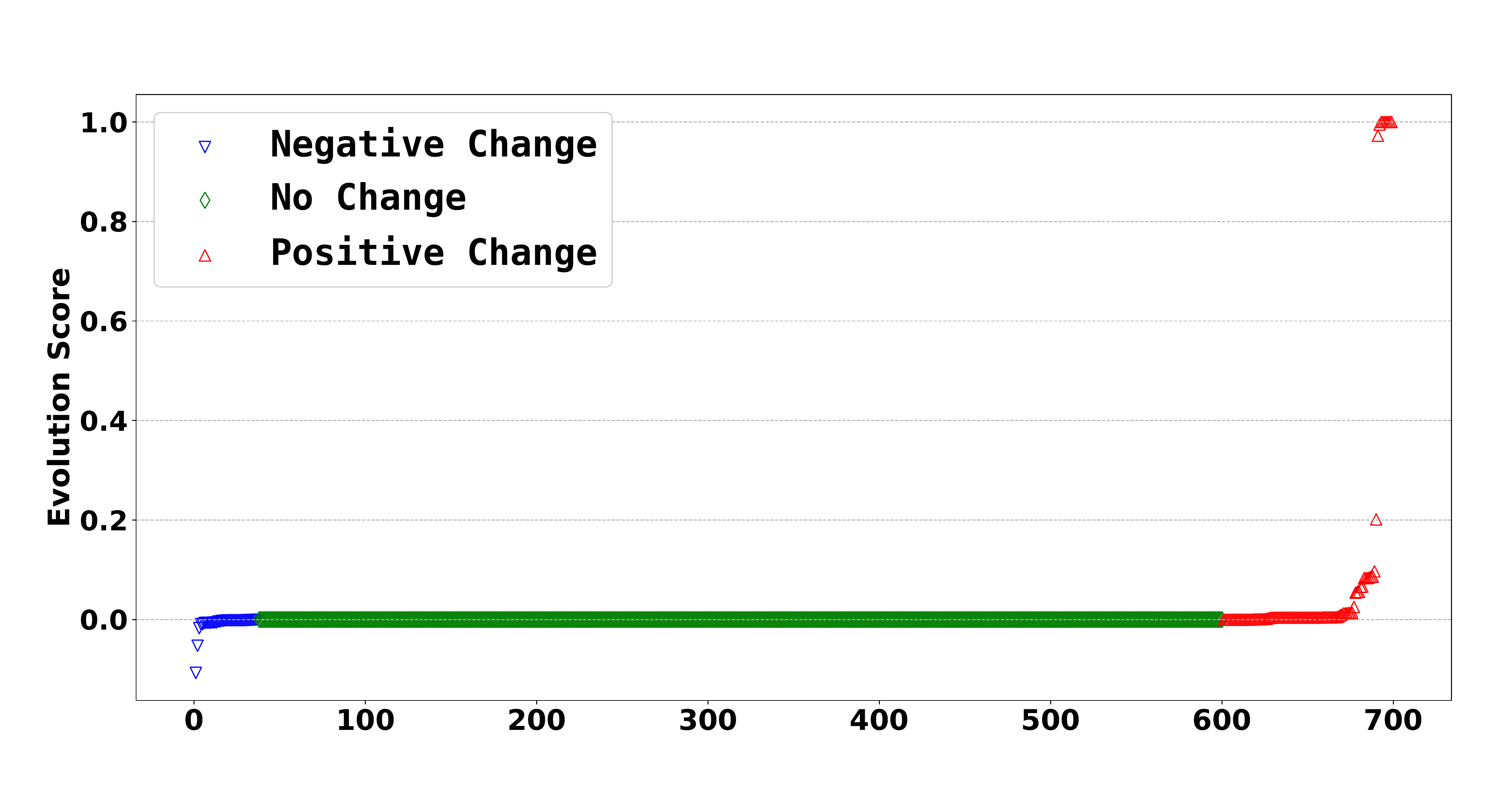}
        \caption{NX Stack}
        \label{fig:product-nx}
    \end{subfigure}
    \begin{subfigure}[b]{0.45\textwidth}
        \centering
        \includegraphics[scale=0.18]{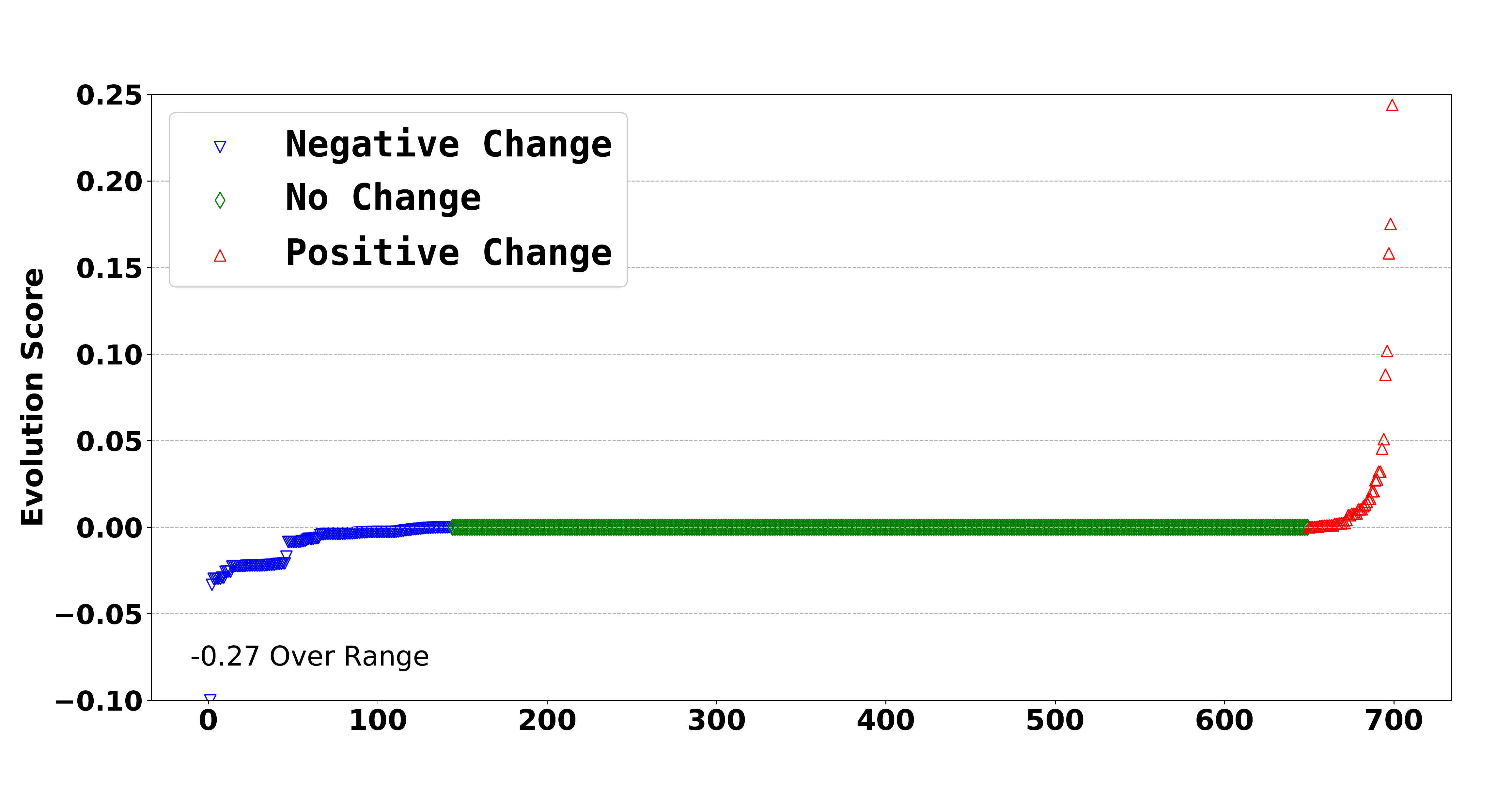}
        \caption{RELRO}
        \label{fig:product-relro}
    \end{subfigure}
    \\
    \begin{subfigure}[b]{0.45\textwidth}
        \centering
        \includegraphics[scale=0.18]{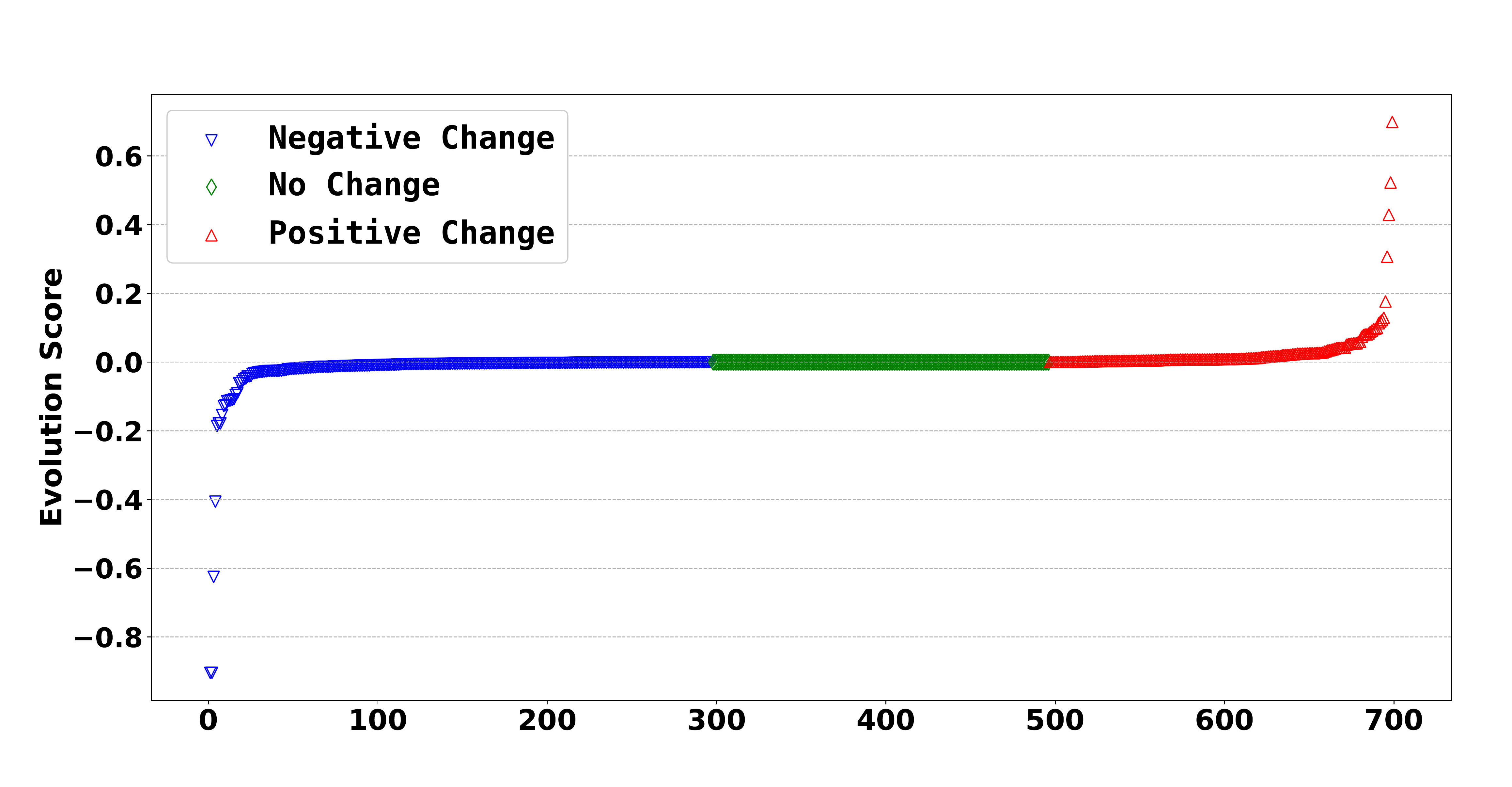}
        \caption{PIE}
        \label{fig:product-pie}
    \end{subfigure}
    \begin{subfigure}[b]{0.45\textwidth}
        \centering
        \includegraphics[scale=0.18]{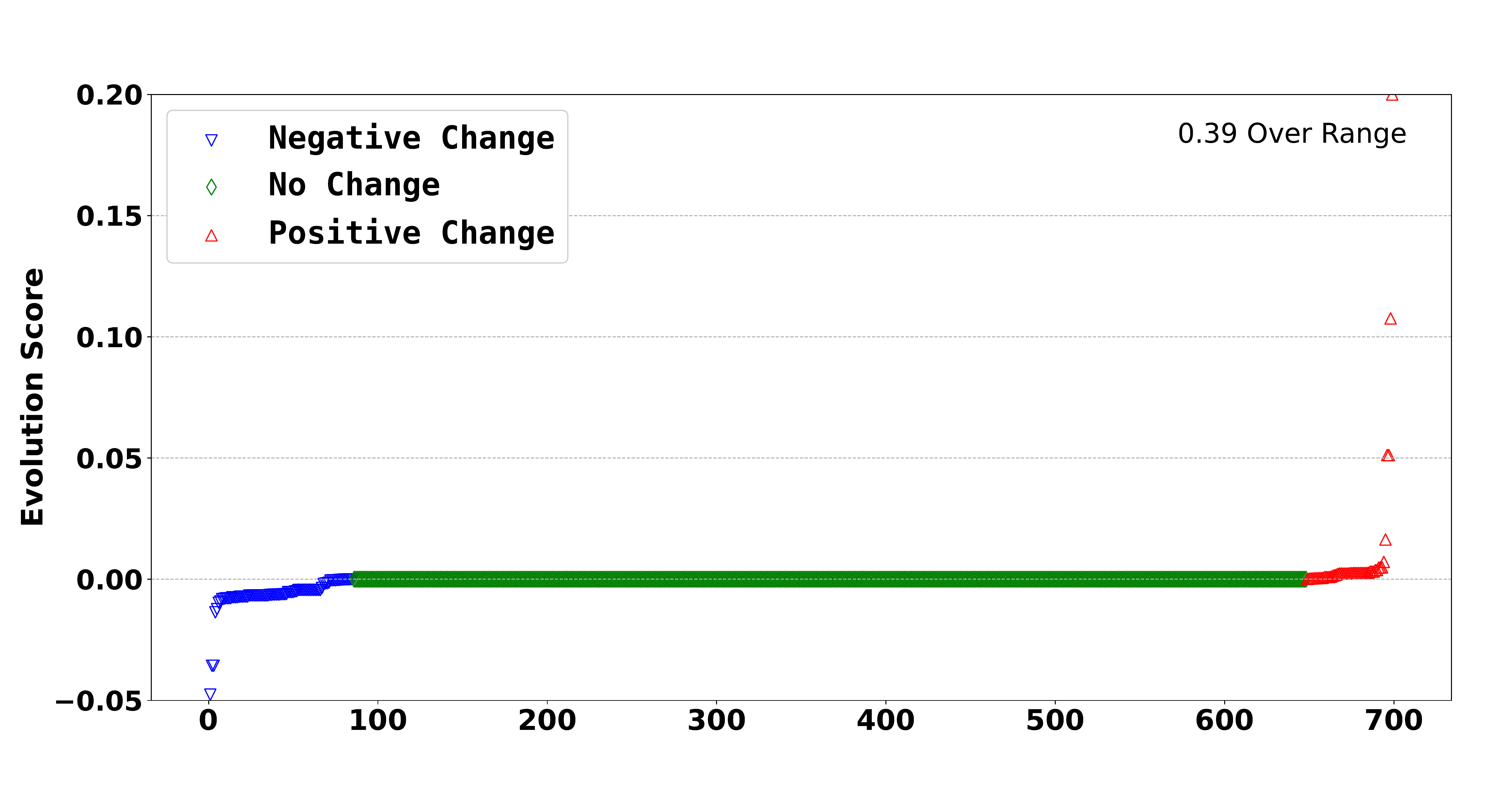}
        \caption{Fortify}
        \label{fig:product-fortify}
    \end{subfigure}
\caption{Evolution score of individual firmware in the adoption of different mitigations. 
        Each point represents a firmware with multiple versions. The firmware 
        is sorted based on the evolution score.}
\label{fig:product-change-other}
\end{figure*}

\setlength\tabcolsep{8pt}
\begin{table*}[t]
  \centering
  \caption{Breakdown results of evolution score of firmware with multiple versions.
  \code{[scr1, scr2]} in each cell means the evolution score for all firmware of the same vendor ranges from \code{scr1} to \code{scr2}.}
  \label{tab:mitigation-vendor-range}
  \footnotesize
  \begin{tabular}{ l | c| c| c| c| c| c}
    \toprule
    \textbf{Vendor} & \textbf{\# of Firmware} & \textbf{Canary} & \textbf{RELRO} &\textbf{NX} & \textbf{Fortify} & \textbf{PIE} \\\toprule
    360 & 1 & [7.4, 7.4] &  [0, 0] & [0, 0]   & [0, 0] & [0, 0]\\ \hline
    AT\&T & 1 & [0, 0] &  [0, 0] &[0, 0] &  [0, 0]  & [0, 0]\\ \hline
    Buffalo & 1 & [0, 0] &  [0, 0]  &[0, 0] & [0, 0] & [0, 0] \\ \hline
    Phicomm & 1 & [0, 0] &  [0, 0] &[0, 0] & [0, 0] & [0, 0]  \\ \hline
    TENVIS & 1 & [0, 0] &   [0, 0]  &[0, 0] & [0, 0] & [0, 0]\\ \hline
    TRENDnet & 2 & [0, 0.3] & [0, 2.8] &[0, 0] & [0, 0] & [0, 2.8]   \\ \hline
    Zyxel & 2 & [0, 0] &  [0, 0]  &[0, 0] & [0, 0] & [0, 0] \\ \hline
    AVM & 3 & [0, 0.3] &    [0, 5.1]  &[-0.06, 0] & [0, 0] &[0, 5.1]\\ \hline
    Moxa & 3 & [0, 10.1] &  [0, 15.8] &[0, 20.1] & [0, 10.8] & [0, 15.8]  \\ \hline
    Mercury & 5 & [0, 0] &  [0, 0] &[0, 0] & [0, 0] & [0, 0]  \\ \hline
    TomatoShibby & 5 & [0, 0.02] &  [0, 0.2] &[0, 5.6] & [0, 0] & [0, 0.2]  \\ \hline
    Netcore & 6 & [0, 0] &  [0, 0] &[0, 0] & [0, 0] & [0, 0]  \\ \hline
    RouterTech & 7 & [0, 0] &  [0, 0] &[0, 0] & [0, 0] & [0, 0]  \\ \hline
    MikroTik & 8 & [-0.01, 0] &  [-0.2, 0]  &[0, 0.01] & [-0.1, 0] & [-0.2, 0] \\ \hline
    Belkin & 12 & [0, 0] &  [0, 3.2] &[-10.7, 100] & [0, 0] & [0, 3.2]  \\ \hline
    QNAP & 12 & [-15.1, -13.8] & [-3.0, -2.2]  &[-0.2, 0.08] & [-0.7, -0.2] & [-3.0, -2.2]  \\ \hline
    Linksys & 14 & [-0.01, 2.4] &  [-0.1, 24.4]  &[0, 99.5] & [0, 0] & [-0.1, 24.4] \\ \hline
    Dlink & 19 & [-3.6, 0.8] &   [0, 0.2] &[-0.4, 100] & [-3.6, 0] & [0, 0.2] \\\hline
    Tenda & 61 & [-0.1, 7.8] & [-0.8, 4.5]  & [0, 0]& [0, 0] & [-0.8, 4.5]  \\ \hline
    Tp-Link-en & 72 & [-0.03, 7.2] &  [-0.3, 0.1]  &[-0.3, 0.09] &[-0.05, 5.1]  & [-0.3, 0.1] \\ \hline
    Tp-Link-zh & 76 & [-7.2, 7.6] & [-1.7, 2.1]  &[-0.7, 100] & [0, 0.7]&  [-1.7, 2.1] \\ \hline
    Synology & 94 & [-3.8, 40.1] &  [-2.3, 17.5]  &[0, 0.5] & [-1.4, 38.6] & [-2.3, 17.5]\\ \hline
    ASUS & 143 & [-2.0, 1.0] & [-2.1, 1.0]  &[-0.9, 9.7] & [-0.8, 0.5] & [-2.1, 1.0]\\ \hline
    Netgear & 150 & [-4.8, 60.3] &  [-27.4, 10.2]  &[-5.3, 8.3] & [-4.8, 1.6]  & [-27.4, 10.2] \\ \bottomrule
    
  \end{tabular}
  \vspace{-1.5em}
\end{table*}

\setlength\tabcolsep{6pt}

\begin{figure}[ht]
        \centering
        \includegraphics[scale=0.19]{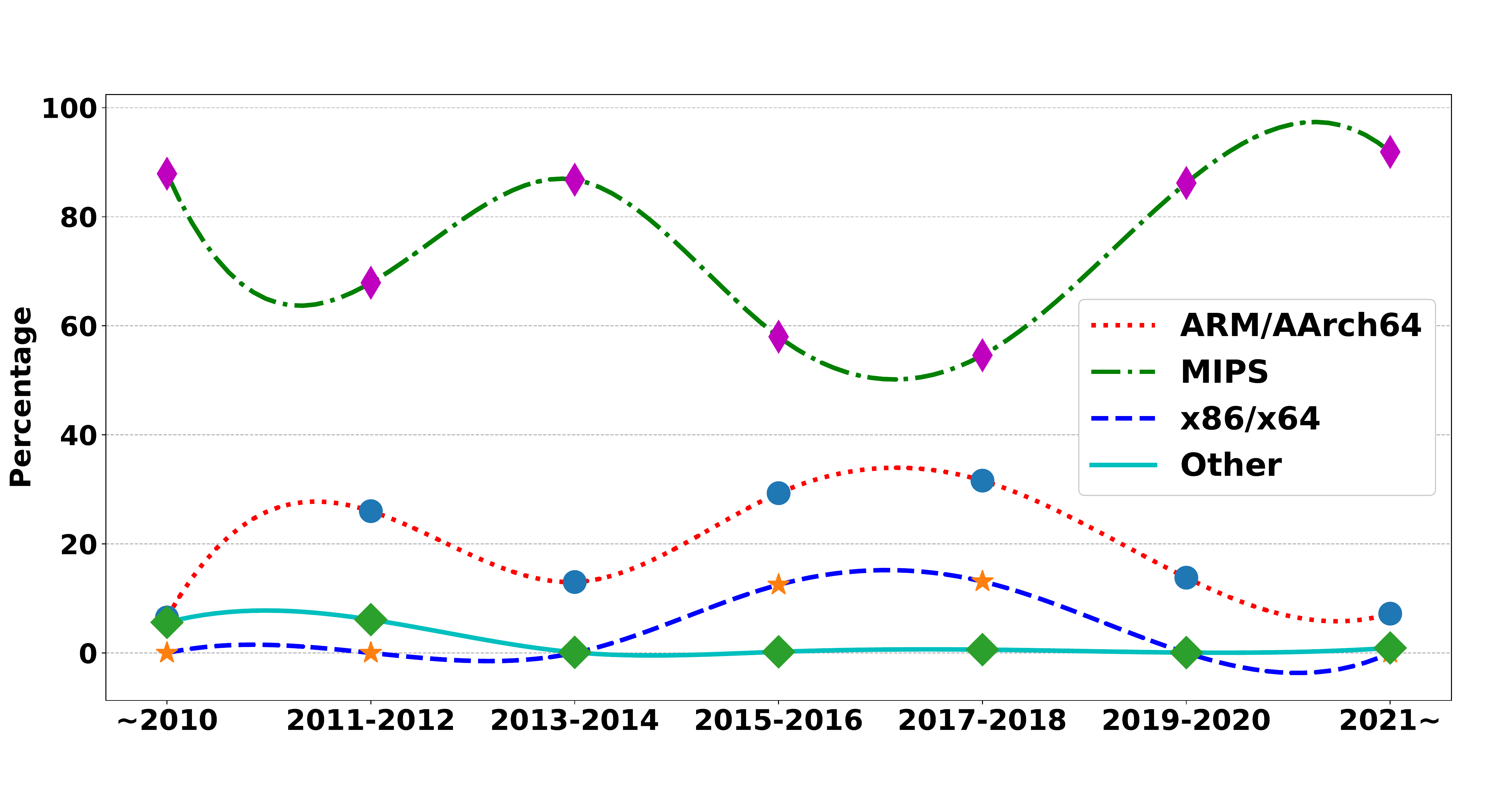}
        \caption{Distribution of binaries based on their architectures. All binaries released 
        before 2010 are aggregated into \code{$\sim$2010}.}
        \label{figure:binary-arch-time}
\end{figure}

\begin{figure}[ht]
        \centering
        \includegraphics[scale=0.44]{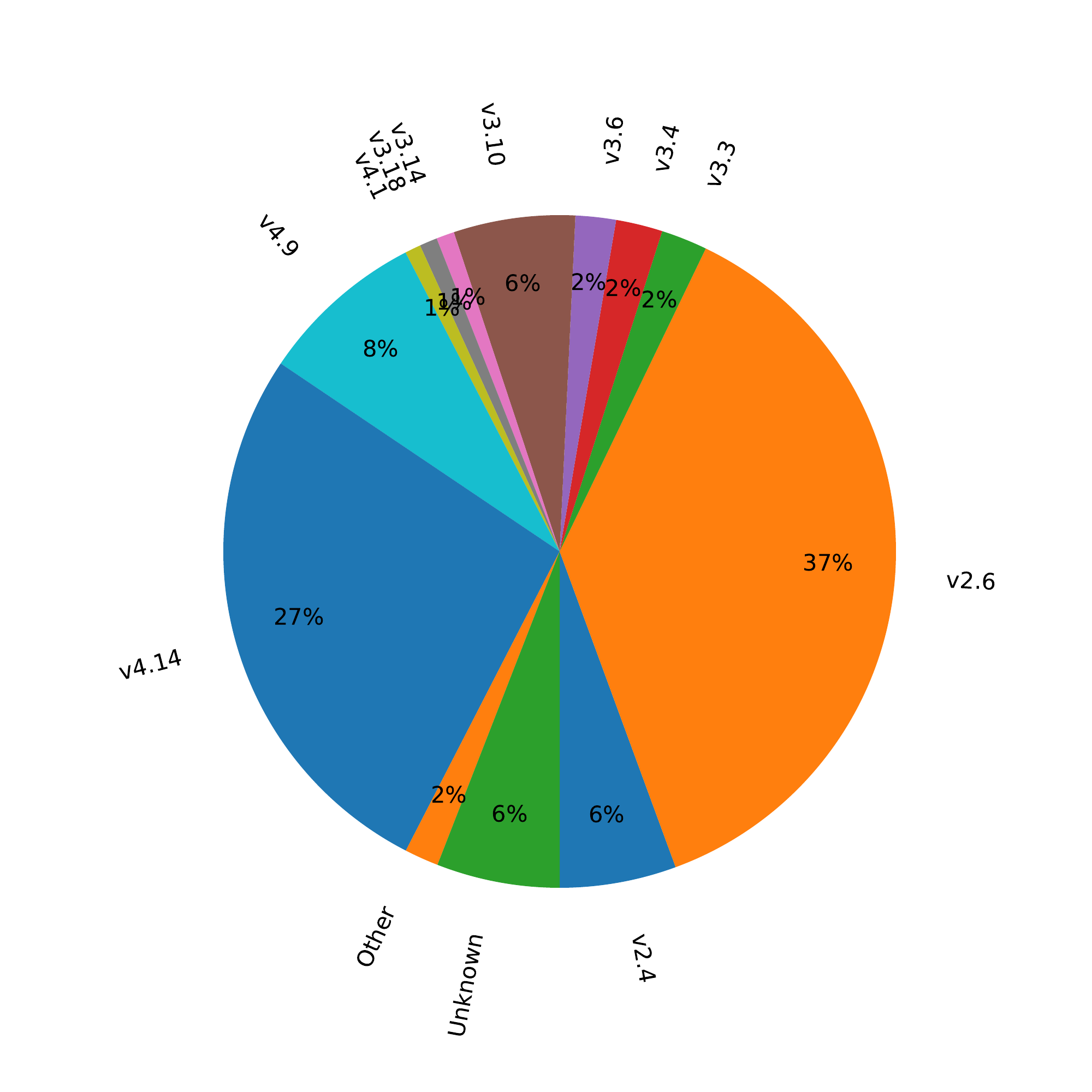}
        \caption{Distribution of different kernel versions.}
        \label{figure:kernel-version-spread}
\end{figure}

\begin{figure}[t]
        \centering
        \includegraphics[scale=0.195]{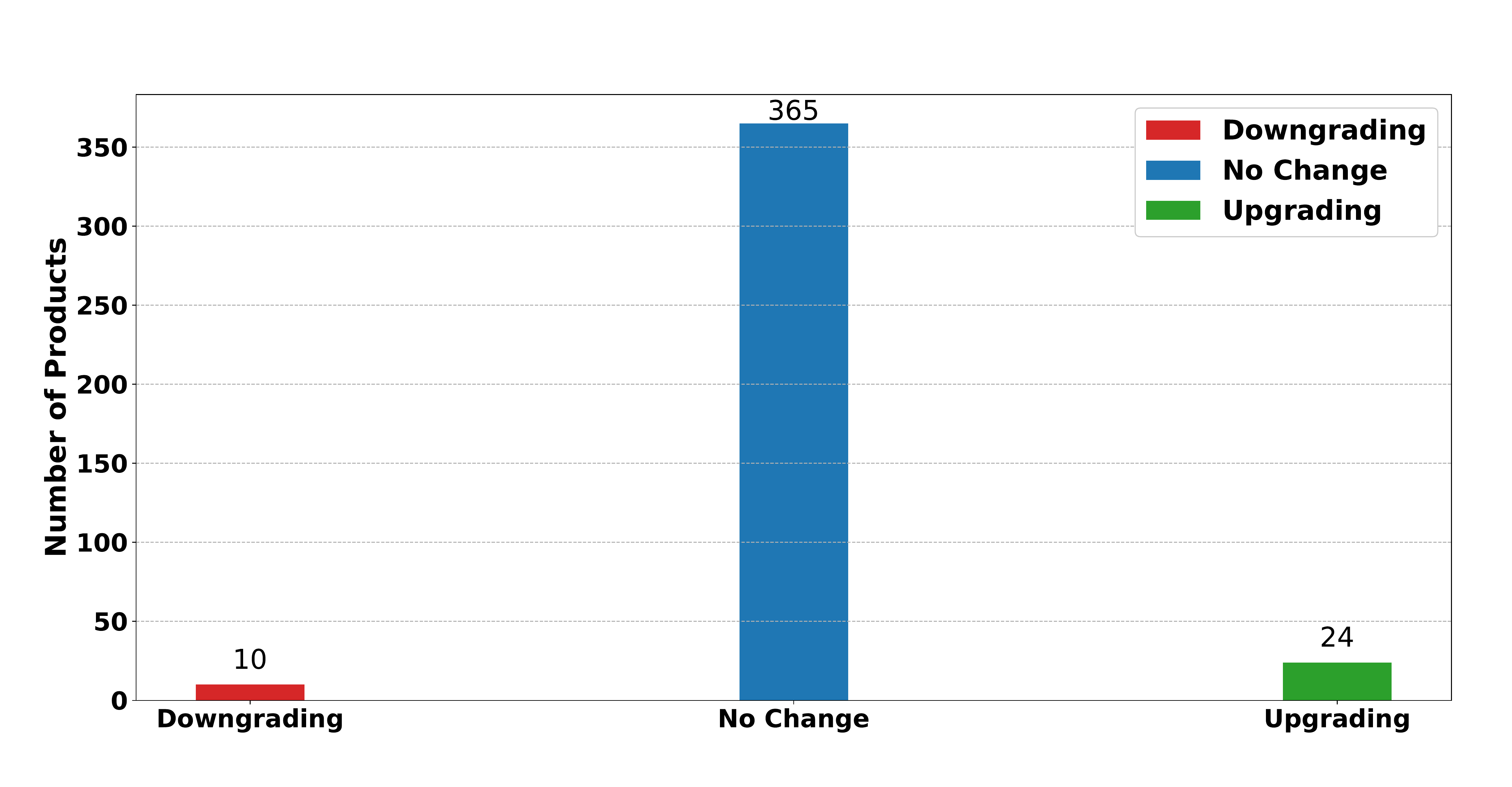}
      \caption{Changes of kernels running on different versions of the same firmware. \code{Downgrading}, \code{No Change}, \code{Upgrading} respectively indicate the number of firmware with kernels downgraded to older versions, with kernels remained at the same version, and with kernels upgraded to newer versions.}
      \label{figure:kernel-version-change}
\end{figure}

\begin{table}[ht]
  \centering
  \caption{Types of device covered in our study}
  \label{tab:fw-category}
    \begin{tabular}{ c }
    \toprule
    \begin{minipage}{0.9\linewidth}
    \raggedright
    {Router},
    {Web Camera},
    {Network Port},
    {Network Switch},
    {Network Storage},
    {Network Access Point},
    {Repeater},
    {Adapter},
    {WIFI Extender},
    {WIFI System},
    {WIFI Bridge},
    {Controller},
    {Video Recorder},
    {Radio},
    {Mother Board},
    {Gateway},
    {Media Connector},
    {Printer},
    {Firewall Modem}
    \end{minipage}\\
    \bottomrule
  \end{tabular}
\end{table}

\setlength\tabcolsep{6pt}
\begin{table}[!t]
  \centering
  \caption{Availability of attack mitigations in different versions of \buildroot.}
  \label{tab:buildroot-mitigation}
  \centering
  \scalebox{0.69}{
  \begin{tabular}{ l | c| c| c| c| c| c}
    \toprule
    \textbf{Version} & \textbf{Default Kernel} & \textbf{Canary} & \textbf{SC$^1$ Dependency} &\textbf{RELRO} & \textbf{Fortify} & \textbf{PIE} \\\toprule
    2021-02 & v5.10 & \checkmark & \checkmark  &  \checkmark  & \checkmark & \checkmark \\ \hline
    2020-11 & v5.4 & \checkmark & \checkmark  &  \checkmark  & \checkmark & \checkmark \\ \hline
    2019-11 & v4.19 & \checkmark & \checkmark  &  \checkmark  & \checkmark & \checkmark \\ \hline
    2018-11 & v4.16 & \checkmark & \checkmark  &  \checkmark  & \checkmark &  \cellcolor{lightgray}$\boxtimes$ \\ \hline
    2017-11 & v4.13 & \checkmark & \checkmark  &   \cellcolor{lightgray}$\boxtimes$  &  \cellcolor{lightgray}$\boxtimes$ &  \cellcolor{lightgray}$\boxtimes$ \\ \hline
    2016-11 & v4.8 & \checkmark & \checkmark  &   \cellcolor{lightgray}$\boxtimes$  &  \cellcolor{lightgray}$\boxtimes$ &  \cellcolor{lightgray}$\boxtimes$ \\ \hline
    2015-11 & v4.3 & \checkmark & \checkmark  & \cellcolor{lightgray}$\boxtimes$  &  \cellcolor{lightgray}$\boxtimes$ &  \cellcolor{lightgray}$\boxtimes$ \\ \hline
    2014-11 & v3.17 & \checkmark & \checkmark  & \cellcolor{lightgray}$\boxtimes$  &  \cellcolor{lightgray}$\boxtimes$ &  \cellcolor{lightgray}$\boxtimes$ \\ \hline
    2013-11 & v3.11 & \checkmark & \checkmark  & \cellcolor{lightgray}$\boxtimes$  &  \cellcolor{lightgray}$\boxtimes$ &  \cellcolor{lightgray}$\boxtimes$ \\\hline
    2012-11 & v3.6 & \checkmark &  \cellcolor{lightgray}$\boxtimes$  &   \cellcolor{lightgray}$\boxtimes$  &  \cellcolor{lightgray}$\boxtimes$ &  \cellcolor{lightgray}$\boxtimes$ \\ \hline
    2011-11 & v3.1 & \checkmark &  \cellcolor{lightgray}$\boxtimes$  &   \cellcolor{lightgray}$\boxtimes$ &  \cellcolor{lightgray}$\boxtimes$ &  \cellcolor{lightgray}$\boxtimes$ \\ \hline
    2010-11 & v2.6 & \checkmark &  \cellcolor{lightgray}$\boxtimes$  &   \cellcolor{lightgray}$\boxtimes$  &  \cellcolor{lightgray}$\boxtimes$ &  \cellcolor{lightgray}$\boxtimes$ \\ \hline
    2009-11 & v2.6 & \checkmark &  \cellcolor{lightgray}$\boxtimes$   &   \cellcolor{lightgray}$\boxtimes$  &  \cellcolor{lightgray}$\boxtimes$ &  \cellcolor{lightgray}$\boxtimes$ \\ \bottomrule
    \multicolumn{7}{ l }{$^{1}$~``SC'' is short for Stack Canaries.}
  \end{tabular}
  }
\end{table}
\setlength\tabcolsep{6pt}

\begin{figure}[!t]
        \centering
        \includegraphics[scale=0.19]{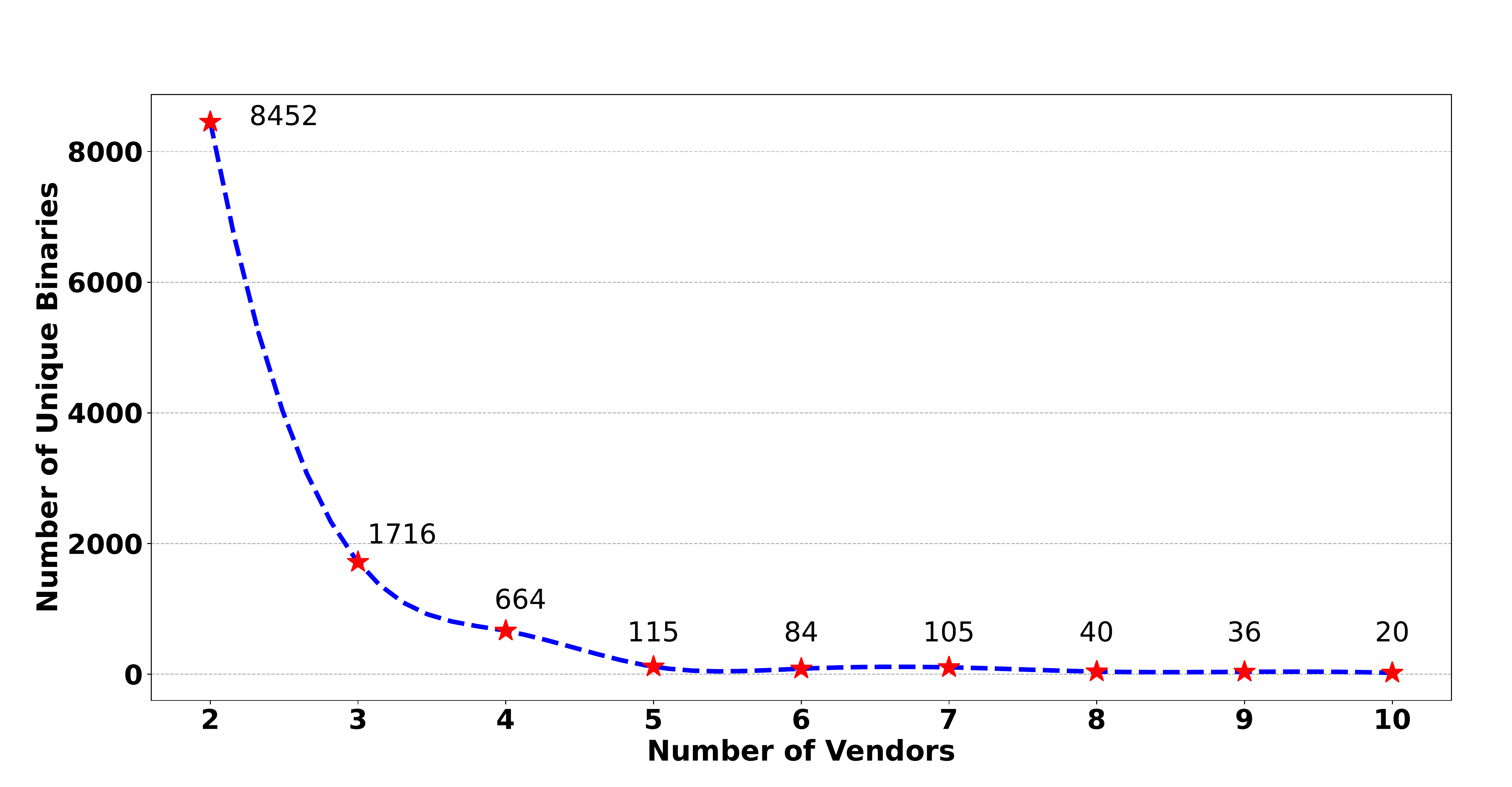}
      \caption{Number of unique binaries reused by multiple vendors. For example, the point at \code{x = 2} means 8,452 unique binaries are reused by 2 vendors (without considering binaries reused by more than 2 vendors).}
      \label{fig:overlap-vendors}
\end{figure}

\setlength\tabcolsep{4.6pt}
\begin{table}[!t]
  \centering
  \caption{Adoption rates of user-space mitigations by binaries shared by multiple vendors (\%).}
  \label{tab:overlap-mitigation-result}
  \begin{tabular}{ c| c| c| c| c| c}
 \toprule
 \textbf{\# of Binaries}  & \textbf{Canary} & \textbf{RELRO}  & \textbf{NX} & \textbf{Fortify} & \textbf{PIE} \\\toprule

11232  & 15.4    & 9.2  & 46.0    & 15.3    & 34.5    \\  \bottomrule
  \end{tabular}
\end{table}

\setlength\tabcolsep{6pt}

\setlength\tabcolsep{8pt}
\begin{table*}[!t]
  \centering
  \caption{Evolution of individual binaries in adopting attack mitigations (breakdown results). In each cell, \code{+x / -y} indicates \code{x} binaries have the mitigation added and \code{y} binaries have the mitigation dropped; ``-'' means no change.}
  \label{tab:mitigation-overlap-binary}
  \footnotesize
  \begin{tabular}{ l | c| c| c| c| c| c}
    \toprule
    \textbf{Vendor} & \textbf{ELF} & \textbf{Canary} & \textbf{RELRO} &\textbf{NX} & \textbf{Fortify} & \textbf{PIE} \\\toprule
    360 & 118 & - & -  &  -  & - & - \\ \hline
    AT\&T & 190 & - &  - & - & -   &  - \\ \hline
    Buffalo & 9 & - & -   & - & - & - \\ \hline
    Phicomm & 147 & - & -  & - &  - &  -  \\ \hline
    TENVIS & 36 & -  &  -    & - &  - & - \\ \hline
    TRENDnet & 409 & +1 / -0 & +6 / -0 & - &  - &  -  \\ \hline
    Zyxel & 154 & -  &  -   & - &  - & -  \\ \hline
    AVM & 358 & +6 / -0 & +20 / -0   & - & -  & +11 / -0 \\ \hline
    Moxa & 107 & - & -  & - & - & -  \\ \hline
    Mercury & 232 &  -  &  -  & - &  - &  -  \\ \hline
    TomatoShibby & 735 & - & - & - &  - &  - \\ \hline
    Netcore & 296 & -  &  -  & - &  - &  -  \\ \hline
    RouterTech & 1086 & -  &  -  & - &  - & +65 / -0   \\ \hline
    MikroTik & 1100 & - & -  & - & -  & - \\ \hline
    Belkin & 688 &  -  & -  & +69 / -0 &  - & -  \\ \hline
    QNAP & 9192 & +11 / -0 & +0 / -39  & - & +26 / -0  & +26 / -0 \\ \hline
    Linksys & 3420 & +1 / -0 & +3 / -0  & +171 / -0 & -  & +4 / -3 \\ \hline
    Dlink & 2399 & - & -  & +236 / -0 & -  & - \\\hline
    Tenda & 7266 & +8 / -0 & -  & - &  - & +17 / -0 \\ \hline
    Tp-Link-en & 9728 & - & -  & - &  - & +0 / -3 \\ \hline
    Tp-Link-zh & 10110 & +2 / -2 & -  & +221 / -0 &  - & +0 / -14 \\ \hline
    Synology & 179949 & +13 / -233 & +177 / -9  & +139 / -0 & +19 / -88  & +353 / -0 \\ \hline
    ASUS & 31581 & +9 / -0 & +0 / -1  & +6 / -1 & -  & +1 / -0 \\ \hline
    Netgear & 20289 & +387 / -49  & +117 / -73  & +164 / -218  & +16 / -16  & +333 / -95 \\ \bottomrule
    
  \end{tabular}
  \vspace{-1.5em}
\end{table*}
\setlength\tabcolsep{6pt}

\begin{figure*}[ht]
        \centering
        \includegraphics[scale=0.43]{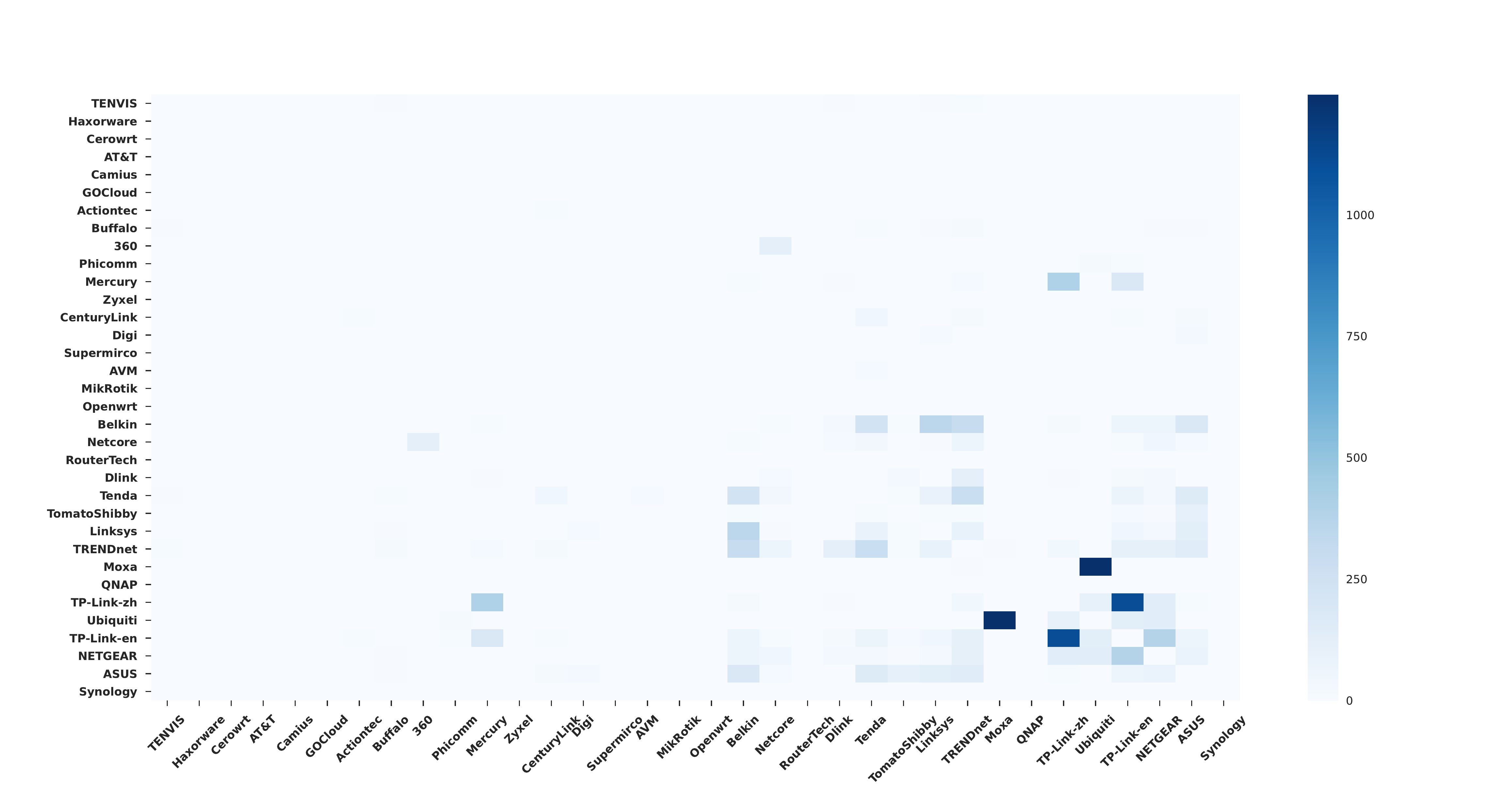}
      \caption{Heatmap showing how many binaries vendors borrow from each other.}
      \label{tab:overlap-heatmap}
\end{figure*}

\end{document}